\documentclass[12pt]{article}
\usepackage{subfig}
\usepackage{graphicx}
\usepackage{graphics}
\usepackage{multirow}
\usepackage{bm}

\begin{document}

\begin{titlepage}
	\vspace*{\stretch{1.0}}
	\begin{center}
		\Large\textbf{Accordion-like metamaterials with tunable ultra-wide low-frequency band gaps}\\
		\large\textit{A O Krushynska$^1$, A Amendola $^2$, F Bosia$^3$, C Daraio$^4$, N M Pugno$^{1,5,6}$ and F Fraternali$^2$}
		
		$^1$ Laboratory of Bio-Inspired and Graphene Nanomechanics, Department of Civil, Environmental and Mechanical Engineering, University of Trento, Via Mesiano, 77,38123 Trento, Italy
		
		\textit{akrushynska@gmail.com, nicola.pugno@unitn.it}
		
		$^2$ Department of Civil Engineering, University of Salerno, Via Giovanni Paolo II, 132, 84084 Fisciano (SA), Italy
		
		\textit{adaamendola1@unisa.it,  f.fraternali@unisa.it}
		
		$^3$ Department of Physics and Nanostrucured Interfaces and Surfaces Centre, University of Turin, Via P. Guria, 1, 10125 Turin, Italy
		
		\textit{fbosia@unito.it}
		
		$^4$ Engineering and Applied Science, California Institute of Technology, Pasadena, CA 91125, USA
		
		\textit{daraio@caltech.edu}
		
		$^5$ School of Engineering and Materials Science, Queen Mary University of London\\ Mile End Road, London E1 4NS, UK
		
		$^6$ Ket Labs, Edoardo Amaldi Foundation, Italian Space Agency\\Via del Politecnico snc, Rome 00133, Italy	
	\end{center}
	\vspace*{\stretch{2.0}}
\end{titlepage}

\vspace{10pt}
%\begin{indented}
%\item[]March 2018
%\end{indented}

\begin{abstract}
We study lightweight, elastic metamaterials consisting of tensegrity-inspired prisms, which present wide, low-frequency band gaps. For their realization, we alternate tensegrity elements with solid discs in periodic arrangements that we call "accordion-like" meta-structures. We show through analytical calculations and numerical simulations that these structures are characterized by low-frequency band gaps with strong uniform wave attenuation due to the coupling of Bragg scattering and local resonance mechanisms. This coupling helps to 
overcome the inherent limit of a narrow band-gap width for conventional locally-resonant metamaterials and to extend the wave attenuation to wider frequency ranges.
Moreover, the band gaps can be further increased, provided a minimum structural damping is present, and tuned to desired frequencies by changing the applied prestress levels in the tensegrity structure. Results are corroborated by parametric studies showing that the band-gap frequencies are preserved for wide variations of geometric and material structural properties. 
\end{abstract}

\noindent{\it Keywords}: wave dynamics, elastic metamaterial, tensegrity structure,  ultra-wide band gap, low-frequency range
%\submitto{\NJP}
%\maketitle

\section{Introduction}\label{intro}

The rapidly developing field of elastic metamaterials opens up promising application opportunities, including  seismic wave shielding~\cite{ktd15, crg16, mk16}, sub-wavelength imaging~\cite{zyj09, md15}, noise and vibration abatement~\cite{sp98, ho2003, mkmbp16}, protective materials~\cite{sd10, aa2014}, acoustic cloaking~\cite{fegm08, zxf11}, just to mention a few examples (see e.g. reviews~\cite{hlr14,lau15,cca16}). 
In general, this is due to their capability of generating band gaps - frequency ranges in which wave propagation is inhibited. Researchers have reported various ways to generate band gaps, such as periodically alternating structural or material properties~\cite{ms95, sp98}, internal resonances~\cite{liu2000, wcswb14, kmbp17}, micro-perforations~\cite{aa93, ho2003, gar12} or 
elongation of wave paths~\cite{ll12, cheng15,kru17}. 
An important challenge is the generation of band gaps at low-frequency scales. A common approach based on incorporation of  resonators, such as inclusions or pillars, leads to bulky structures and a limited band-gap width with non-uniform Fano-type profile for wave attenuation~\cite{liu2000, kkg14, coffy15, luyang17, md17}.
To maximize the gap width, researchers have developed rainbow-trapping designs~\cite{ktd15}, coupling of the Bragg and local resonant mechanisms~\cite{coffy15}, and elaborated powerful topology optimization techniques~\cite{bh11,laude2017}, as well as multi-scale and multi-objective  methods~\cite{hh06, aj16}. It has also been proposed to remove certain parts of the material from a unit cell (i.e., introduce cavities) that reduces the effective material stiffness, and thus, lowers the Bragg scattering limit~\cite{md17, dac16}.

%%-------------------------------- 
\begin{figure}
	\centering
	\includegraphics[scale=0.8]{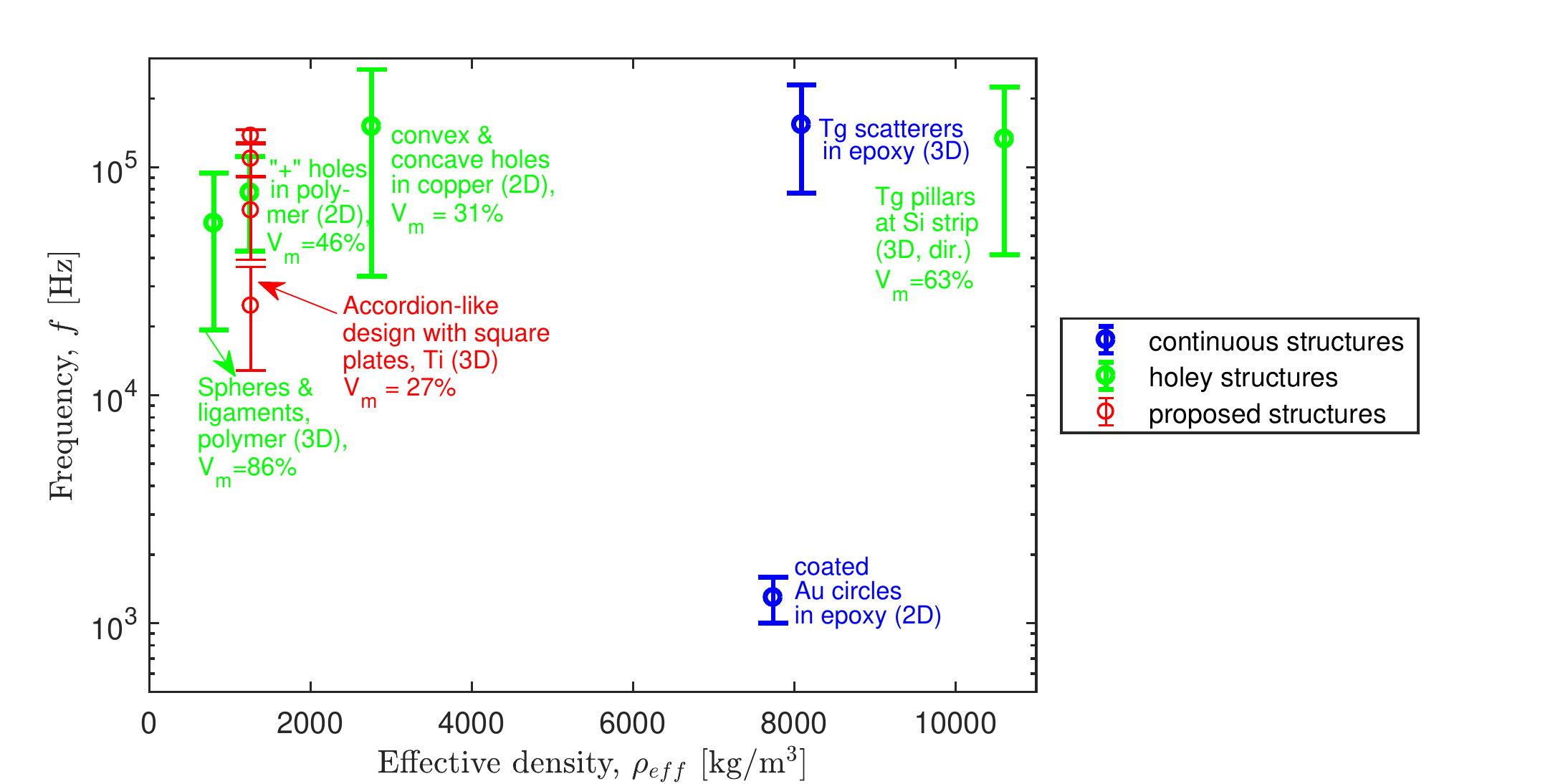}
	\caption{Band-gap widths for elastic metamaterials with unit cell size $h_{uc}=10$mm. $V_m$ denotes the material volume fraction. The chosen configurations represent metamaterials with local resonators, pillars or cavities and optimized configurations with wide low-frequency band gaps.
		The data are taken from Ref.~\cite{kkg14} for coated Au circles in epoxy; Ref.~\cite{laude2017} for convex \& concave holes in copper; Ref.~\cite{wang11} for cross-like holes in polymer; Ref.~\cite{dac16} for polymeric spheres joint by ligaments; Ref.~\cite{luyang17} Tg scatterers in epoxy; Ref.~\cite{coffy15} from Si strip with Tg pillars; from Section~\ref{sec:3Dmodel} for accordion-like configurations.
		\label{fig_BGsize}}
\end{figure}

Figure~\ref{fig_BGsize} shows band-gap widths for several metamaterial configurations with local resonators, pillars or cavities, as well as some optimized configurations with wide low-frequency band gaps. These include three-dimensional and two-dimensional (in-plane wave polarization) structures.
Bold lines correspond to continuous configurations without air/vacuum cavities, while thin lines correspond to holey structures.
The horizontal axis indicates the values of  effective metamaterial density $\rho_{eff}$ evaluated as a sum of densities for each material phase multiplied by the material volume fraction. 
The band-gap widths
have been evaluated for a unit cell size of 10~mm, based on the data provided in the original works~
\cite{kkg14,laude2017,wang11,dac16,luyang17,coffy15}, to make the comparison fair.

The data in Fig.~\ref{fig_BGsize} provides selected examples of approaches for the generation of low-frequency band gaps. One of them is based on the incorporation of heavy resonators and results in dramatical increase of the structural weight~\cite{coffy15, luyang17}. By tuning the resonators packing density and increasing its weight, one can reach very low-frequency ranges~\cite{kkg14}.
Another approach consists of the introduction of cavities and a proper redistribution of  weight and effective stiffness for the remaining structural parts in order to maximize the band-gap width. 
Thus, recently found meta-structures with slender elements exhibit wide band gaps, while preserving light weight~\cite{dac16, laude2017,md17}. However, for single-phase configurations, it appears to be difficult to lower the band-gap frequencies, as can be seen from the data in Fig.~\ref{fig_BGsize}. 
Moreover, for many practical applications, it is of importance to enable tunability of band-gap frequencies. 
The reported tuning strategies include harnessing instabilities~\cite{bvwfb15,wcswb14} or thermal radiation~\cite{wrrkwn2014}, exploitation of  piezoelectric effects~\cite{cdber12} or magnetic nonlinearities~\cite{bfd17}, and  incorporation of rotational elements~\cite{lnflhc11,glgt13}, etc. These strategies require either specific non-trivial material behavior or/and complicated setups implying additional constraints on the metamaterial topology. Such requirements can be incompatible with optimized designs or drastically increase exploitation costs.
Therefore, the quest continues for simple structured lightweight metamaterials with wide low-frequency band gaps, which can be actively tuned in operating conditions.

In this work, we propose 
a new design strategy for lightweight 
metamaterials that can strongly attenuate low-frequency waves 
within only a few unit cells. The designed topologies exploit advantages of tensegrity structures (e.g., light-weight and rich design possibilities), and allow  to easily tune pass and stop bands by varying the level of applied prestress. 

The proposed meta-structures are composed of tensegrity lattices attached to bulk discs~\cite{fsd12}. 
Typical tensegrity elements consist of bars connected by cables. Their shape and mechanical performance are governed by tensile stress in cables applied on either the initial configuration (`local' prestress) or during operation (`global' prestress)~\cite{ow00, so10, fra14, dmp16}.  
Such a peculiar organization enables to achieve a variety of mechanical behaviors ranging from extreme stiffening to extreme softening~\cite{fra14, fca2015}, which can be effectively tuned by adjusting the prestress level, as demonstrated experimentally~\cite{aa2014}. Tensegrity lattices alternating lumped discs have been shown to support  tunable solitary waves with anomalous wave transmission and reflection from interfaces between branches with different acoustic impedances~\cite{fra14}. 

In a theoretical model, we assume that the  tensegrity lattices are rigidly attached to the disks with the continuity condition at the joints. The metamaterial units are then  characterized by a finite stiffness even in the absence of prestress in the tensegrity elements. Thus, the designed structures do not rigorously obey the tensegrity  principle, implying the stabilization of compressed members through pre-stretchable cables, which only bear compressive and tensile stresses, in the absence of bending stresses~\cite{so10}.  We therefore refer to such structures as "accordion-like" metamaterials, due to the resemblance of the overall system combining tensegrity units and bulk masses with the structure of this musical instrument (see Fig.~\ref{fig:unit}a,b).
We focus on the study of linear elastic waves in these configurations and on the numerical analysis of their dispersion and transmission characteristics. We assume an either rigid or elastic behavior for interlaying discs and demonstrate that in both cases the meta-structures are capable of generating wide low-frequency band gaps.
We evaluate the influence of the metamaterial geometric and mechanical parameters on the gap width and 
estimate the band-gap tunability depending on the level of initial prestress in the strings. Finally, we demonstrate that the band gaps are partially preserved for 
a wide range of incident angles for a three-dimensional geometry.

The organization of the paper is as follows. Section~\ref{sec_model} describes the two developed accordion-like metamaterial models. 
In Section~\ref{sec:meta-chain} we analyze wave dynamics in an accordion-like meta-chain under assumption on rigid or elastic behavior of interlaid discs. The effects of the applied prestress on the band-gap tunability are also studied.
In Section~\ref{sec:param_studies}, we perform an extensive analysis of the dependence of wave dispersion characteristics on geometric and material parameters. In Section~\ref{sec:3Dmodel}, we consider a three-dimensional metamaterial model and discuss its wave attenuation abilities depending on the angle of incident waves. Finally, in Section~\ref{sec:concl} we draw the main conclusions.

\section{Accordion-like metamaterial models}\label{sec_model}
\begin{figure}
	%\centering%
	\begin{tabular}{cc}
		\begin{tabular}{c}
			\subfloat[]{\includegraphics[scale=0.35]{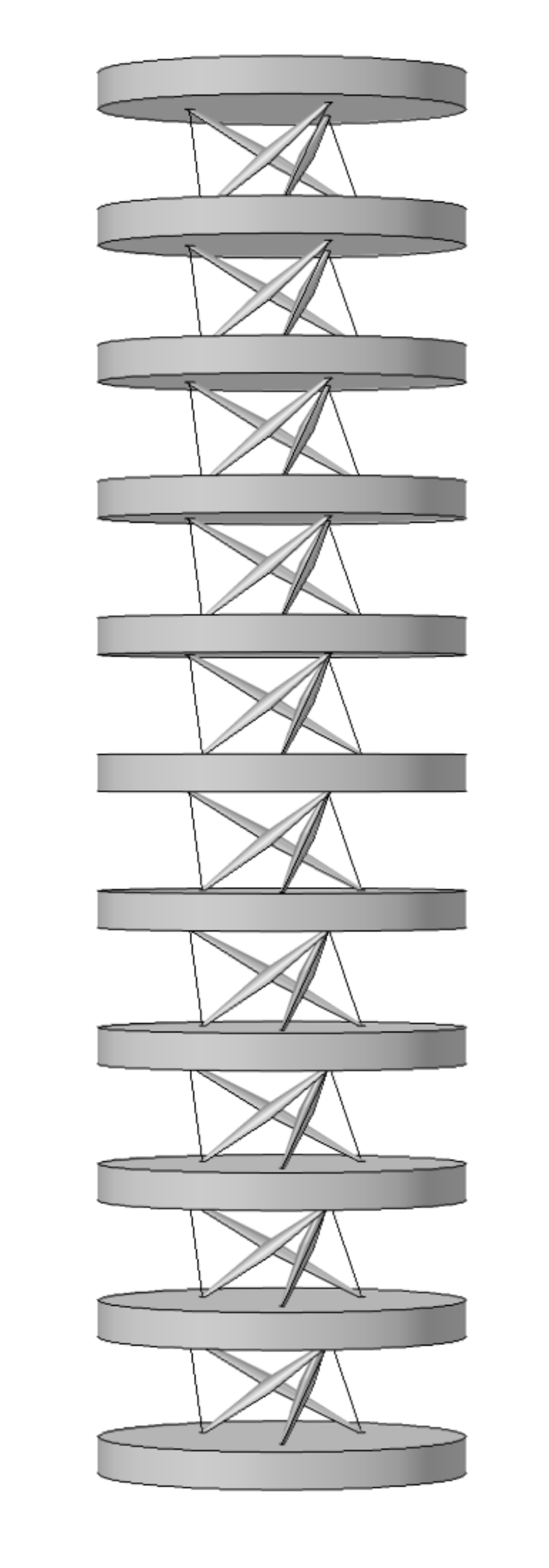}}
		\end{tabular}
		&
		\begin{tabular}{c}
		\subfloat[]{\includegraphics[scale=0.25]{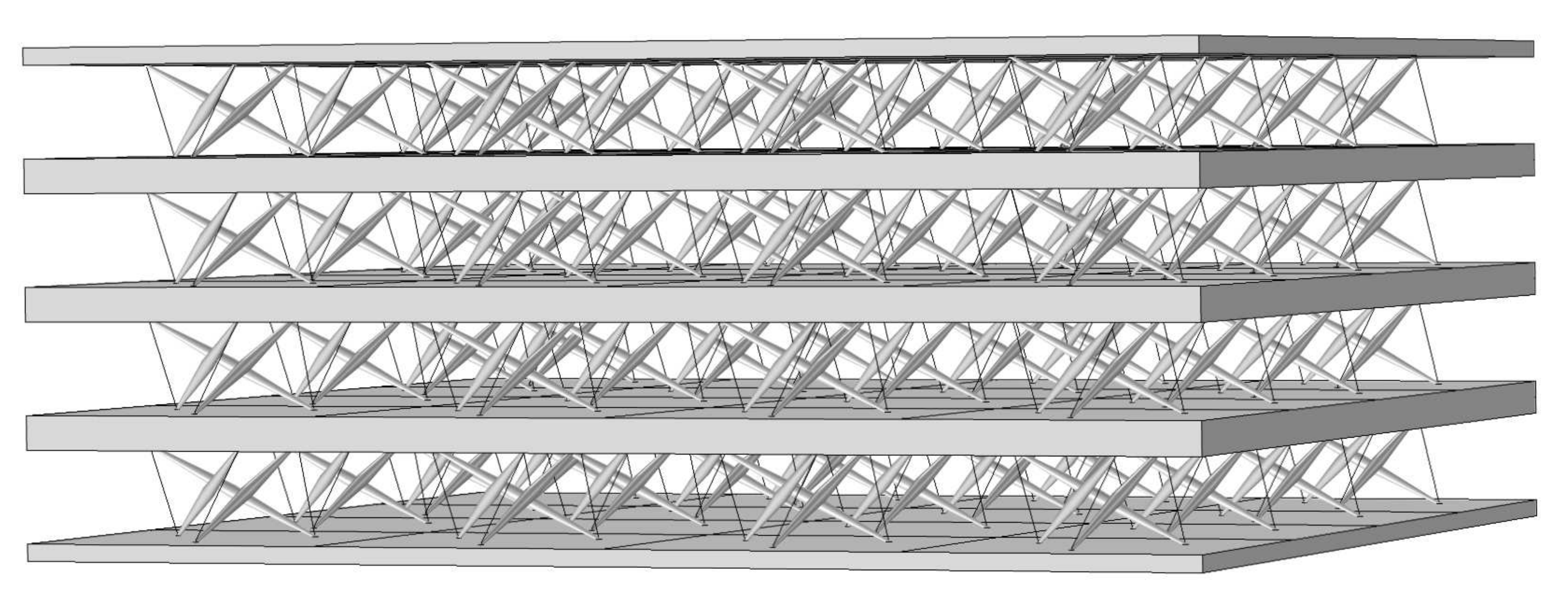}}\\
		\subfloat[]{\includegraphics[scale=0.25]{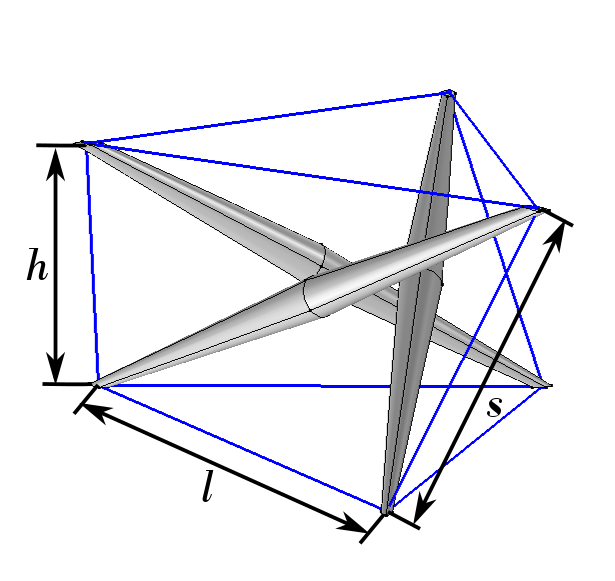}}
		\end{tabular}
	\end{tabular}
	\caption{Accordion-like meta-chain (a) and three-dimensional (b) metamaterial model composed of periodically repeated tensegrity prisms interlaid with bulk discs or square elements; (c) composition of a minimal regular tensegrity prism.}\label{fig:unit}
\end{figure}

We analyze the propagation of small-amplitude linear elastic waves in accordion-like metamaterials in the form of one-dimensional (1-D) chain (Fig.~\ref{fig:unit}a) or a three-dimensional (3-D) structure (Fig.~\ref{fig:unit}b). The two models consists of periodic repetitions of tensegrity prisms interlaid by circular discs or square elements. 
We employ a regular minimal tensegrity prism (Fig.~\ref{fig:unit}c), as defined in \cite{so10}, which is rigidly attached to the terminal discs. There is continuity of displacements at the connection points. 
These structures can be manufactured by means of widely-used technical processes, such as precision-water-jet cutting, additive manufacturing, 3-D printing, assembling of metallic structures with subsequent high-temperature brazing, etc.

%---------------------------------
%---------------------------------

The tensegrity prism is composed by three inclined tapered bars connected by five pre-stressed fibers (or strings).
Two fibers form horizontal equilateral triangles at the ends of the bars. 
We denote the prism height as $h$. The terminal triangles with side length $l$ can be rotated relative 
to each other by an arbitrary twist angle $\phi$. Simple geometric considerations allow to establish a relation between the bar length $b$ and the mentioned parameters
\begin{equation}
b=\sqrt{h^2 + \frac{4}{3}l^2\sin^2\frac{\phi}{2}},
\end{equation}
as well as with the length of cross strings $s$:
\begin{equation}
s=\sqrt{b^2 + \frac{\sqrt{3}}{2}l^2\cos\left(\phi+\frac{\pi}{6}\right)}.
\end{equation}
Note that the assumption on continuity of displacements at the prism-disc joints eliminates the necessity to introduce horizontal strings, which are thus no longer considered.

The strings are assumed to be made of PowerPro\textregistered Spectra fibers of 0.28 mm diameter with Young's modulus $E_f = 5.48$ GPa and mass density $\rho_f$=793 kg/m$^3$~\cite{aa2015}. 
The bars and the discs are assumed of titanium alloy Ti6A14V with Young's modulus $E_t = 120$ GPa, Poisson's ratio $\nu_t=0.33$, and mass density $\rho_t=4450$ kg/m$^3$~\cite{aa2015}. 
The central and end diameters of a bar are $D=0.8$~mm and  $d=0.18$ mm, respectively. In the absence of prestress in the strings, i.e. $p_0=0$, the prism height is $h_0=$5.407 mm and the triangle side is $l_0$=8.7 mm. The wave dynamics of tensegrity metamaterials with other geometric parameters is studied in Section~\ref{sec:param_studies}.

The circular discs and square elements have equal mass and thickness to facilitate the comparison of wave characteristics for the two models. We choose the disc thickness to be $t=2$~mm. 
If the radius of the circular disc is fixed to $R=10$~mm, then the lateral size of the square element is $a=17.72$~mm.
The chosen material and geometric properties result in the same value of the effective density, $\rho_{eff}=1216$ kg/m$^3$, for the both unit cells.

\section{Elastic waves in a tensegrity meta-chain}\label{sec:meta-chain}

\subsection{Rigid-elastic model of a  meta-chain}
\label{sec:rigid-elastic}

We first approximate the discs as rigid elements, neglecting their deformations~\cite{fsd12}. The dynamics of such a disc is then described by the motion of its center of mass.  
For the specified material and geometric parameters, the mass of the tensegrity prism, $m_{p} = 0.034$~g, can be neglected as compared to the mass of a disc, $m_d = 2.8$~g ($m_d/m_p=82.2$).
This allows us to consider the meta-chain, in a first approximation, as a set of point masses connected by non-linear massless springs. Wave propagation in such a chain can be analyzed analytically by means of a 1D mass-spring model~\cite{hlr14, aa2018}.

A nonlinear stress-strain relation for the tensegrity lattice governed by geometric effects has been extensively studied, e.g. in Refs.~\cite{aa2014, fca2015, fsd12}. However, when considering small oscillations of tensegrity elements around their initial positions, as in case of small-amplitude waves, 
one can linearize the system response. As a result, the prism axial stiffness $k_h$ is assumed to be constant and equals the slope of tangent of the stress-strain curve~\cite{aa2018}. 

The dispersion relation for a 1D linear mass-spring system has a single acoustic branch originating from zero and extending up to the cut-off~\cite{bri46, hlr14}:
\begin{equation}
f_{lb}^{a}=\frac{\sqrt{k_{h}/m_d}}{\pi},
\label{Bloch_eq}
\end{equation} 
where the superscript \emph{a} indicates the analytical origin of the bound estimation and the subscript $lb$ designates the lower band gap bound. 
When a wave propagates along the meta-chain, 
it induces variations of the prism height  (translational degree of freedom) and of a twist angle of the discs (twisting rotations)~\cite{fra14, dmp16}. 
Hence, we define the mode described by relation~(\ref{Bloch_eq}) as `translational-twisting'. Note that this mode is analogous to a fundamental compressional wave in a mass-spring chain.

Relation~(\ref{Bloch_eq}) neglects  a  coupling between translational-twisting and other possible modes. In practice, this can be implemented for frictionless contact between the discs and prisms. The axial stiffness $k_h$ of the prism is then governed solely by prestress in the cross-strings and can be evaluated analytically~\cite{fca2015}. In this case, the metamaterial obeys the tensegrity principle, and its dynamics is studied in Ref.~\cite{aa2018}.

%=============================================
\begin{table}[h]
	\caption{Relation between applied static loading $F$ and induced axial displacement $u_z$ in the tensegrity prism. The axial stiffness of the prism is evaluated as $k_h = f/u_z$; frequency $f_{lb}^a$ is estimated by means of Eq.~(\ref{Bloch_eq}). }
	\centering
	\begin{tabular}{cccc}
		$F$~[N]&$u_z$ [m]&$k_h$ [N/m]&$f_{lb}^a$ [Hz]
		\\ \hline
		3.18$\times 10^{-5}$&7.8209$\times 10^{-9}$&4061&384\\
		3.18$\times 10^{-4}$&7.8208$\times 10^{-8}$&4061&384\\
		3.18$\times 10^{-3}$&7.8208$\times 10^{-7}$&4061&384\\ 
		1.59$\times 10^{-2}$&3.9068$\times 10^{-6}$&4065&384\\
		\hline
	\end{tabular}
	\label{tab-stiffness}
\end{table}
%================================================

If the tensegrity lattices are rigidly joint to the disks, the axial stiffness $k_h$ depends, in addition, on 
the tangential stiffness of inclined bars. 
We numerically evaluate the stiffness of a tensegrity prism by means of finite-element simulations in Comsol Multiphysics 5.2~\cite{com52} taking into account possible geometric non-linearities.
In the absence of initial prestress, $p_0=0$, a stable equilibrium configuration is considered, in which 
adjacent discs are rotated relative each other by angle $\phi=5\pi/6$~\cite{ow00, fsd12}.  
A distributed tensile force is applied to a  surface of a top disc of the unit cell, while a bottom disc is clamped. To model rigid behavior, we multiply the Young's modulus $E_t$ by 10$^3$. Calculations demonstrate that further increase of $E_t$ provides similar results. 

The dependence of the axial stiffness $k_h$ on the applied force is illustrated in Table~\ref{tab-stiffness}.
Note that for displacements of order $10^{-6}$, which correspond to small-amplitude excitations, the axial stiffness is constant. This justifies the introduced linearization of the metamaterial response.
%------------------------------------------------
\begin{figure} 
	\begin{tabular}{cc}
		\begin{tabular}{cc}
			\includegraphics[height=12cm]{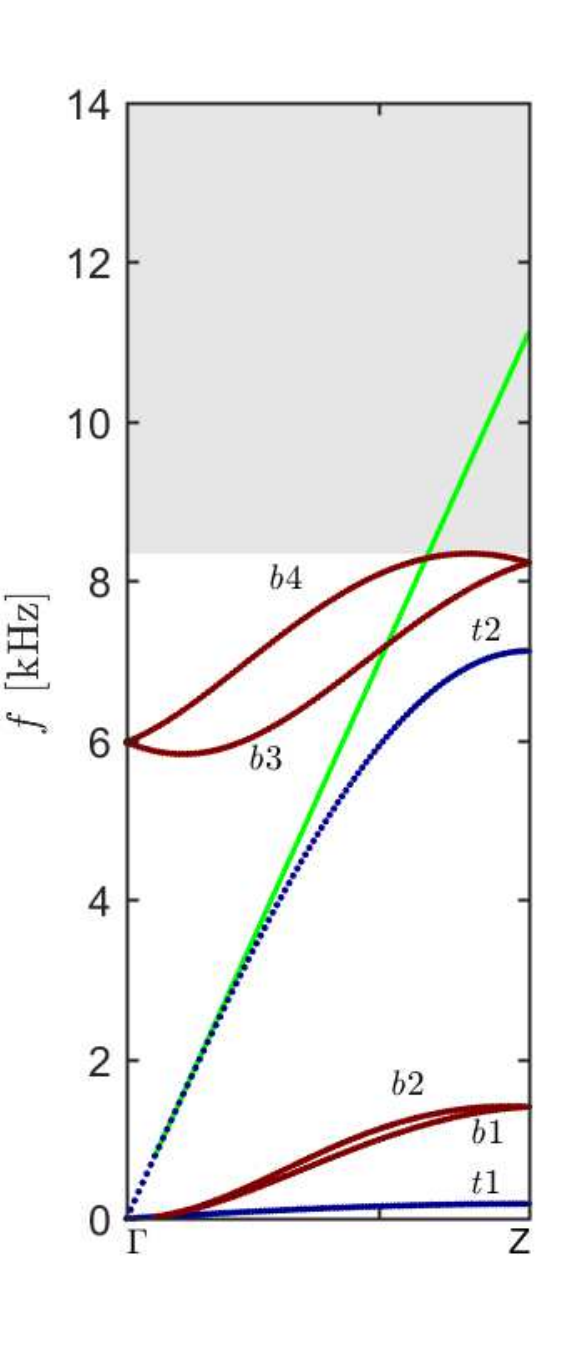}
			&
			\includegraphics[height=12cm]{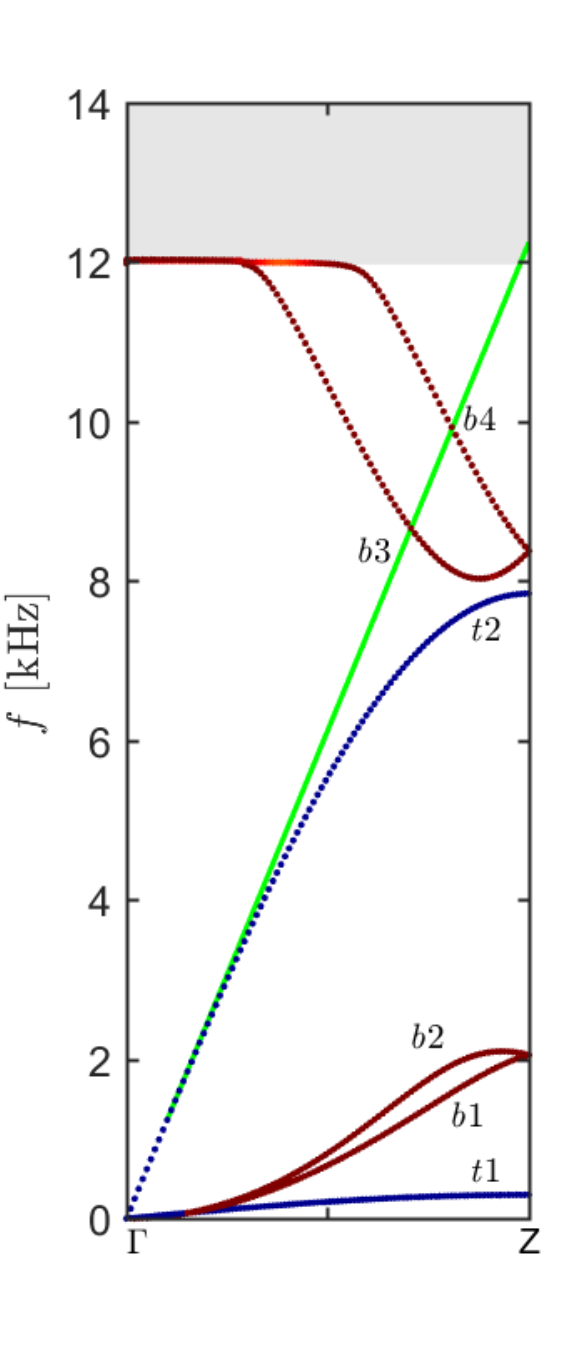}\\
			(a)&(b)\\
		\end{tabular}
		&
		\begin{tabular}{c}
			\includegraphics[scale=0.12]{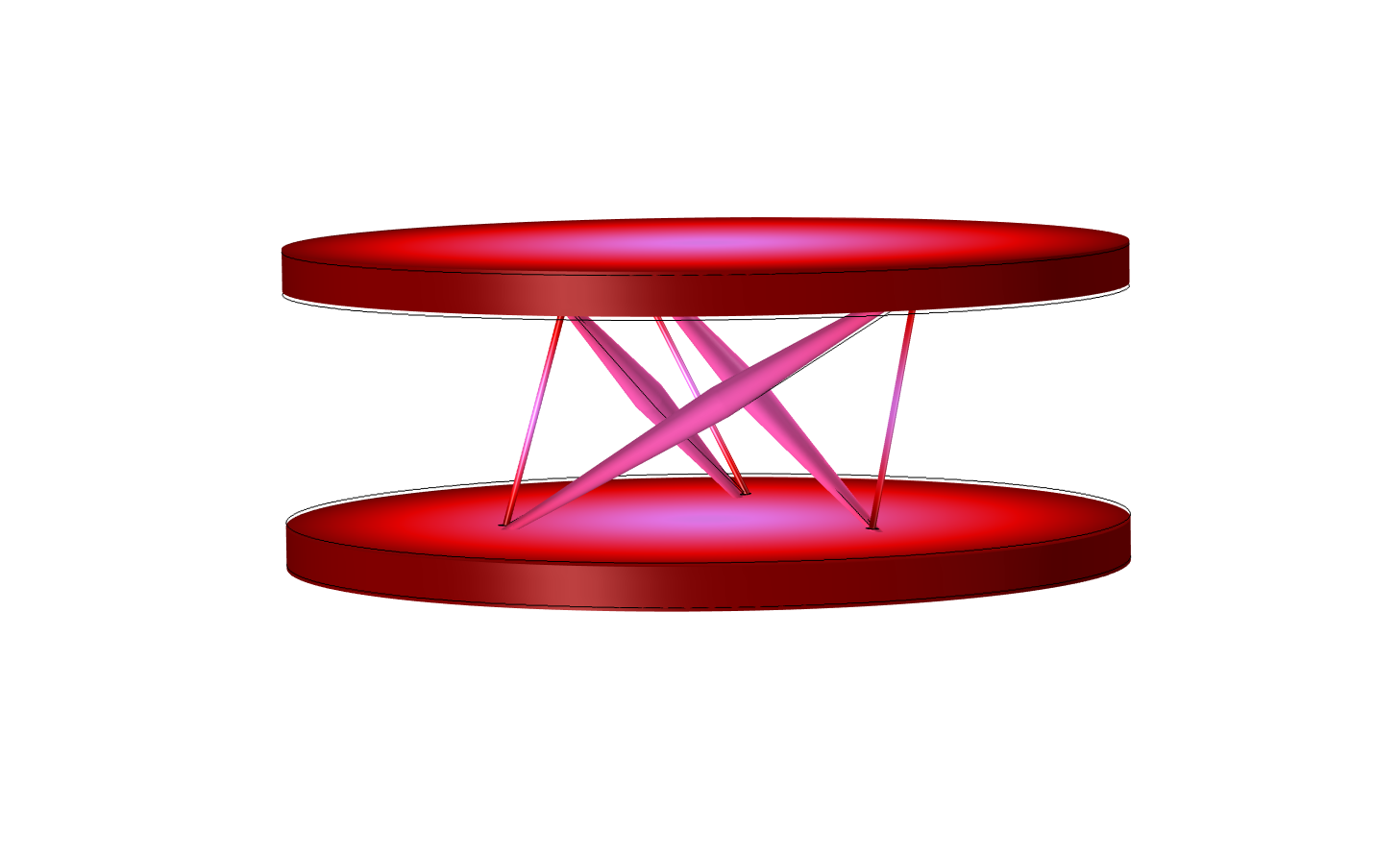}\\
			(c) mode t1\\
			\includegraphics[scale=0.12]{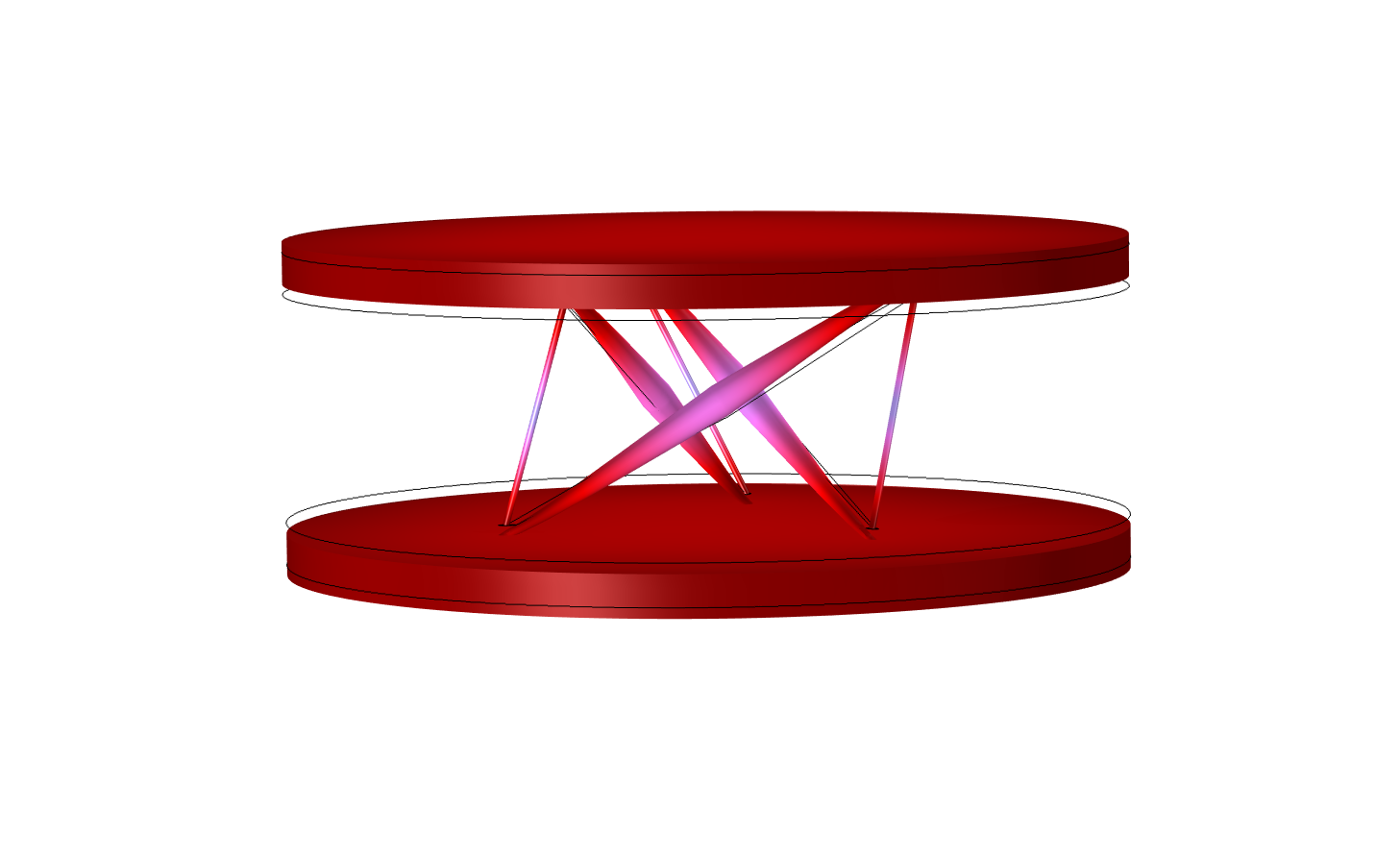}\\
			(d) mode t2\\
			\includegraphics[scale=0.12]{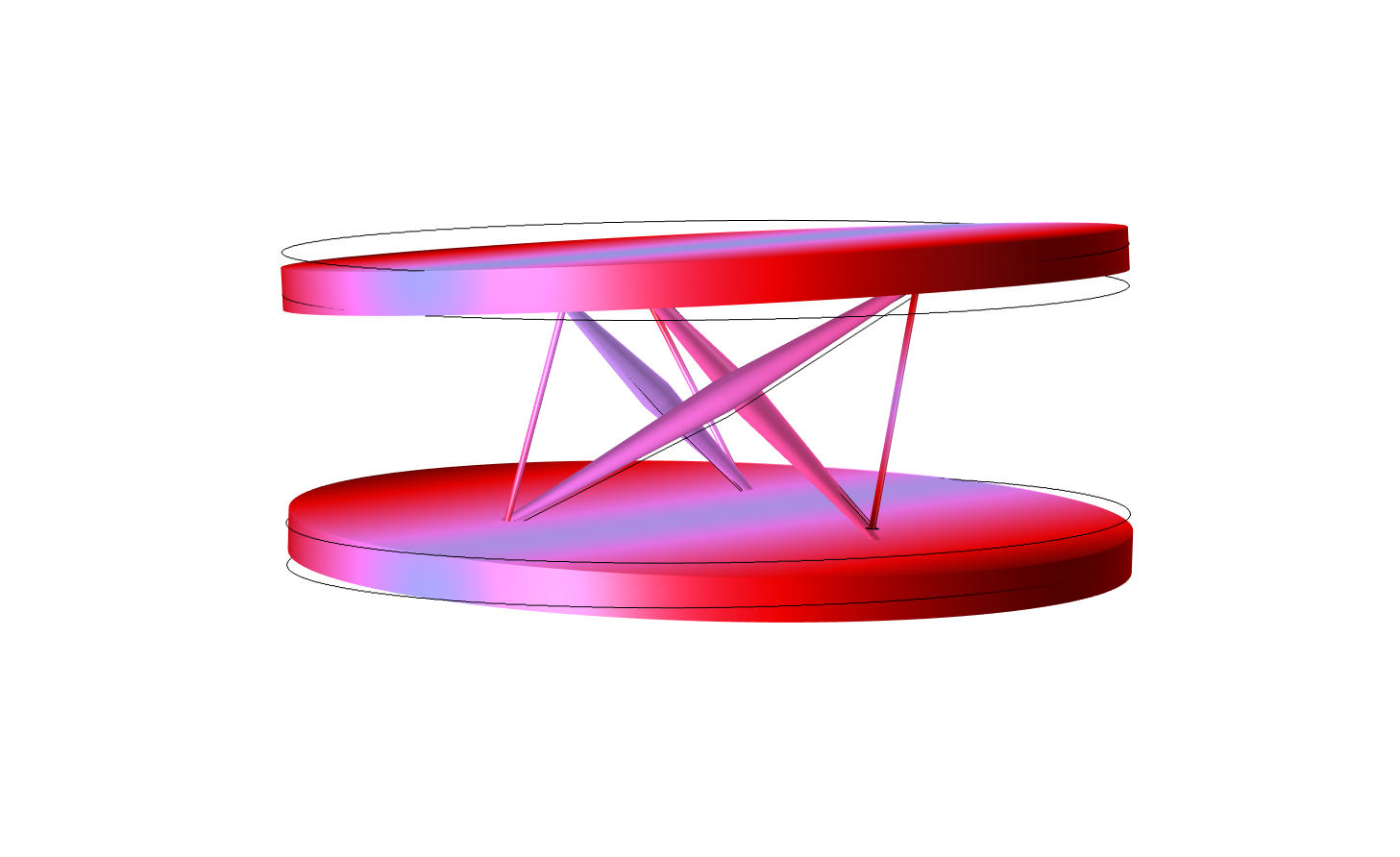}\\
			(e) modes b1 and b2\\
			\includegraphics[scale=0.12]{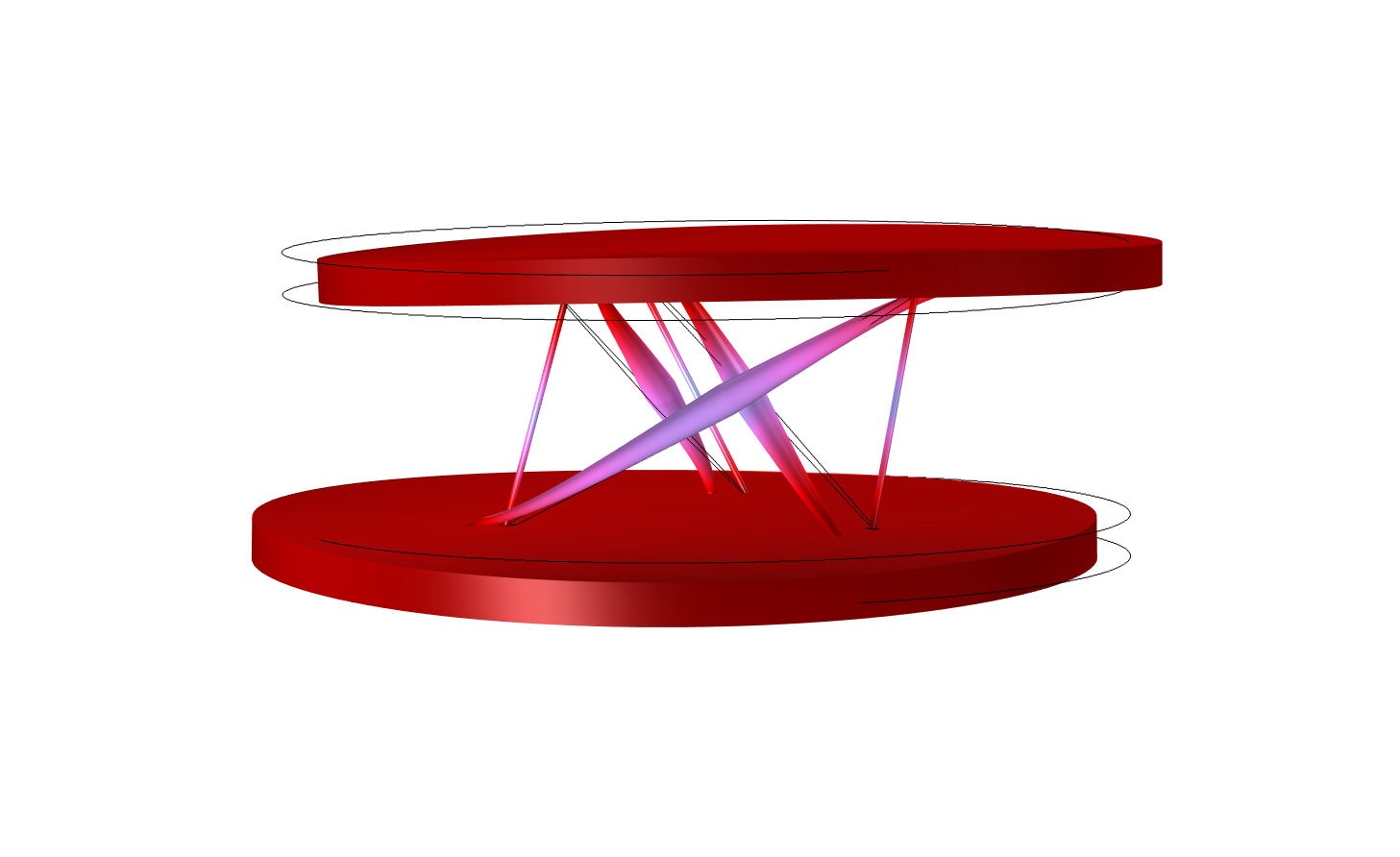}\\
			(f) modes b3 and b4\\
		\end{tabular}
	\end{tabular}
	\caption{{\bf Rigid-elastic model}: Dispersion relations for accordion-like meta-chains with the disk parameters (a) $R=10$~mm, $t=2$~mm and (b) $R=5.6$~mm, $t=6.38$~mm providing identical disk masses. The notations $t$ and $b$ refer to translational-twisting (blue) and bending (red) modes; the number indicates the mode order. Green solid lines are tangents to the $t2$ mode, which intersect lines $k=Z$ at (approximate) mid-gap Bragg frequency. Shaded regions denote band gaps. Mode shapes of translational-twisting (c-d) and bending (e-f) modes at point $Z$ of the Brillouin zone. Red and metallic blue colors in the mode shapes correspond to maximum and zero displacements, respectively.}
	\label{fig:RigEl_spectrum}
\end{figure}
%------------------------------------------------
The low value of stiffness suggests that one can expect the generation of a band gap at extremely low frequencies starting from $f^{a}_{lb}$.

To verify this assumption, we numerically evaluate the dispersion relation for a rigid-elastic metamaterial unit cell.
For this purpose, we apply the Floquet-Bloch conditions at central cross-sections of the discs and consider positive real wavenumbers $k_z$ along the border of the irreducible Brillouin zone $\Gamma-Z$, which corresponds to the axial direction in real space. 
The simulation results are shown Fig.~\ref{fig:RigEl_spectrum}a and reveal the presence of the second $t2$ mode at higher frequencies represented by a blue line. Due to this mode, the band gap is shifted to about 40 times higher frequencies.  The mode shape of the $t1$  mode is characterized by parallel axial motions of the discs and uniformly distributed displacements along the bars of a tensegrity prism (Fig.~\ref{fig:RigEl_spectrum}c). For the $t2$ mode, the bars vibrate non-uniformly with maximum displacements at the joints (Fig.~\ref{fig:RigEl_spectrum}d). Therefore, the second mode appears due to the displacement continuity conditions between the tensegrity prism and the discs. In the both cases, the mode shapes resemble those for typical compressional waves in mass-spring systems.

Note that the numerical prediction for a cut-off of the $t1$ mode, i.e.,  $f_1=184$~Hz, is much lower compared to the analytical estimation $f_{lb}^{a} = 384$ Hz. This discrepancy is attributed to the fact that the analytical solution neglects  geometric dimensions of the discs, which are essential for numerical simulations. To demonstrate this, we evaluated the dispersion relation for an equivalent unit cell with  thicker ($D=3.19$~mm) and smaller ($R=5.6$~mm) discs of the same mass, $m_d=const$ (see Fig.~\ref{fig:RigEl_spectrum}b). For this case, the cut-off of  $t1$ mode, $f_1 = 297$ Hz, approaches the value of $f_{lb}^a$, while the frequency of $t2$ mode slightly changes, as can be expected from its dynamics.
% $f_2 = 7125$ Hz for R=10mm
% $f_2 = 7841$ Hz for R=5.6mm

If the discs are allowed to rotate, the dispersion relations have additional modes that involve bending rotations relative to the disk plane. Hence, we refer to them as `bending' modes.  
Two pairs of bending modes are shown by red lines in Figs.~\ref{fig:RigEl_spectrum}a-b and  labeled as $b1$, $b2$ and $b3$, $b4$, respectively with the mode shapes given in Fig.~\ref{fig:RigEl_spectrum}e-f. The upper pair of rotational modes shifts the band gap bound to slightly higher frequencies.

To shed light on the  band-gap generation mechanism, we estimate a mid-gap Bragg frequency $f_{mid}^{B} = c_p/2h_{uc}$ governed by the structural periodicity, where $c$ is the effective phase velocity in the medium. For our structure with slender elements, the effective phase velocity can be evaluated directly from the dispersion relation as $c_p=2\pi{}f/k$ in the vicinity of $\Gamma$ point for each admissible mode type~\cite{md17}. For translational-twisting modes, frequency $f_{mid}^{B}$ is found at the intersection of a tangent to the $t2$ mode (green solid line in Fig.~\ref{fig:RigEl_spectrum}a,b) and the vertical line $k=Z$. 
As can be seen in Fig.~\ref{fig:RigEl_spectrum}a-b, the mid-gap Bragg frequencies are located inside the band gaps estimated numerically. Hence, the wave attenuation within the band gaps occurs due to the Bragg scattering effect.

%% R_min case
%% trans1: f_1 = 297.58 Hz
%% rotat1: f_2 = 2051.3 Hz
%% rotat2: f_3 = 2051.3 Hz
%% trans2: f_4 = 7840.6 Hz
%% rotat3: f_5 = 8373.3 Hz
%% rotat4: f_6 = 8377.7 Hz
%% localized in bar1: 12015 Hz

%% R = 10 mm case
%% trans1: f_1 = 184.08 Hz
%% rotat1: f_2 = 1401.7 Hz
%% rotat2: f_3 = 1401.8 Hz
%% trans2: f_4 = 7121.4 Hz
%% rotat3: f_5 = 8218.3 Hz
%% rotat4: f_6 = 8237.2 Hz
%% localized in bar1: 11929 Hz

To evaluate the attenuation efficiency, we perform a numerical transmission analysis in the frequency domain. We consider a finite meta-chain composed of 30 unit cells. One end of the chain is excited with a prescribed axial displacement of amplitude $u_{z0}=0.1$~micron applied to the whole disc, while the other end is attached to a perfectly matched layer (of 5 unit cells) to eliminate undesired wave reflections from the boundary.  In order to avoid unrealistically large displacements at resonant frequencies, we introduce a vibration loss factor in the material of tensegrity bars. The loss factor  $\eta$ takes into account inherent structural damping in a dynamically loaded material and enters the stress-strain relationship as
$
\boldsymbol{\sigma} = D\left(1+j\eta\right)\boldsymbol{\epsilon}
$.
We choose $\eta=0.001$ Pa$\cdot$s that corresponds to the minimum value of losses in titanium alloys, as assessed experimentally~\cite{lll08}.  

%----------------------------
\begin{figure}[htb]
	\includegraphics[scale=0.8]{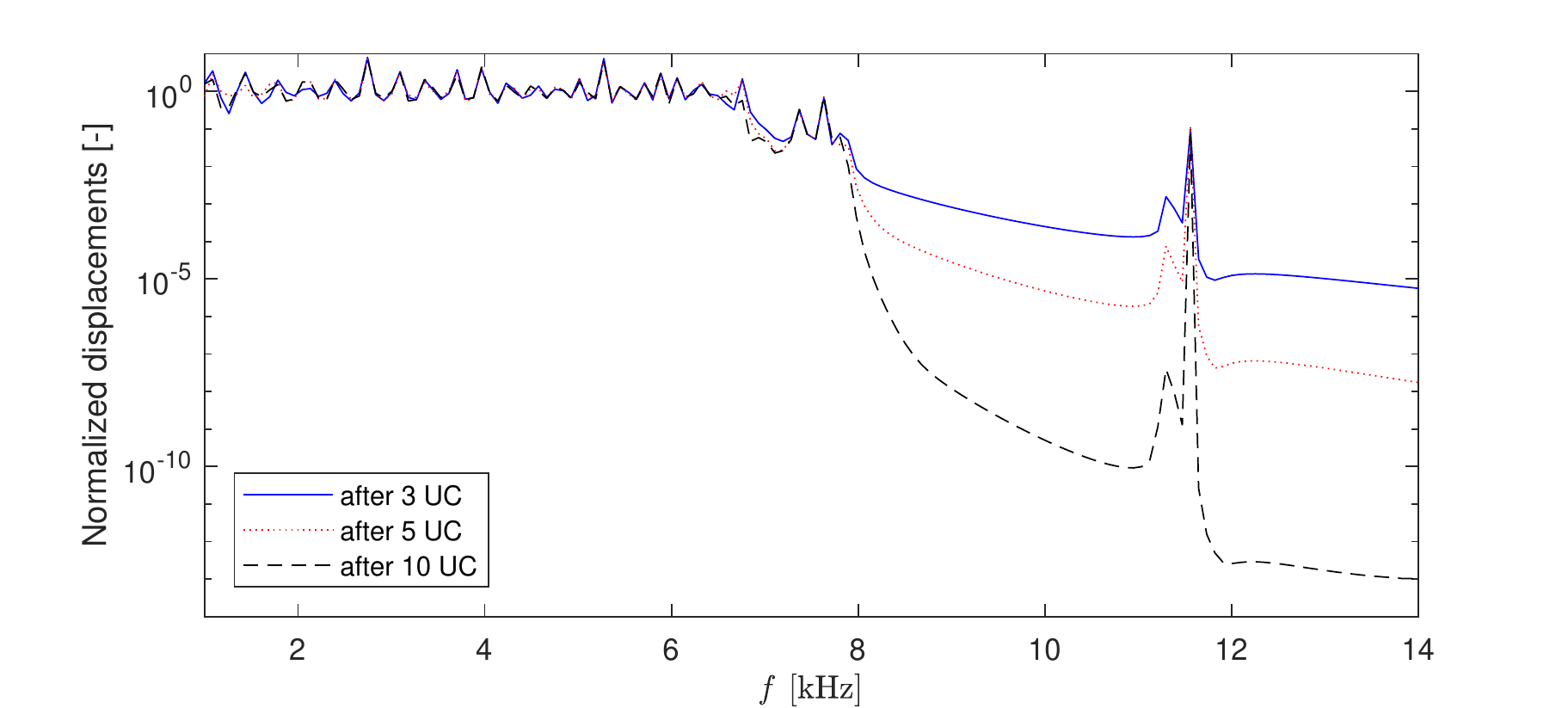}
	\caption{{\bf Rigid-elastic model:} Normalized transmitted displacements $\sqrt{u_x^2+u_y^2+u_z^2}/u_{z0}$ versus frequency $f$ for an accordion-like  meta-chain composed of 30 unit cells (disc dimensions are $R=10$~mm, $t=2$~mm). The curves indicate displacement values averaged on three adjacent discs, which are located at distances of 3,5, or 10 unit cells from the excited end.}
	\label{fig_trans_rigid}
\end{figure}
%----------------------------

Figure~\ref{fig_trans_rigid} shows the magnitude of displacements  $u_{tot}=\sqrt{u_x^2+u_y^2+u_z^2}$ averaged upon three adjacent discs and normalized with respect to the loading $u_{z0}$. The blue, red and black curves correspond to transmitted displacements at distances of 3, 5, and 10 unit cells from the loaded end, respectively. By comparing data in Fig.~\ref{fig_trans_rigid} and the dispersion relation in Fig.~\ref{fig:RigEl_spectrum}a, one can clearly see that below the cut-off frequency of the $t2$ mode, waves propagate along the chain without attenuation. Between the cut-off of the $t2$ mode and the lower band gap bound $f_{lb}=8237$~Hz, the wave amplitude slightly decreases. This occurs due to excitation of rotational modes $b3$ and $b4$ (see the discussion below). Finally, at frequencies higher than $f_{lb}$, the wave attenuation is significant and reaches 5 orders of magnitude for waves passing through only 5 unit cells. The observed strong uniform wave attenuation is typical for the Bragg scattering mechanism~\cite{md17}. 
Several transmission peaks at around 11.5 kHz correspond to localized modes with displacements concentrated in the tensegrity bars (these modes are not shown in Fig.~\ref{fig:RigEl_spectrum}a and are analyzed in Section~\ref{sec: elastic case}). 

The
calculations reveal that despite the symmetry of excitation conditions, the bending modes are present in pass bands. This can be explained by the asymmetric structure of the tensegrity prisms, in which one end is twisted relative to the other. 
Since wave propagation along the meta-chain is accompanied by a simultaneous change of the twist angle $\phi$ and the prism height $h$, the non-symmetric location of joints at adjacent discs makes rotational motion unavoidable even for perfectly symmetric, irrotational loading.	
Hence, the designed meta-chains with rigid discs and continuous displacements at the joints cannot support propagation of translational-twisting waves only, in contrast to tensegrity meta-chains with frictionless contact conditions~\cite{aa2018}.

\subsection{Elastic model of a meta-chain}
\label{sec: elastic case}

In this section, we take into account deformations in the disks and assume elastic behavior for all the constituents of an accordion-like meta-chain. 
First, we assume that an initial prestress in the cross-strings is absent, i.e., $p_0=0$, and the tensegrity unit is in its equilibrium configuration with $\phi=5\pi/6$~\cite{fra14}. 

\begin{figure}[htbp] 
	\begin{tabular}{cc}
		\begin{tabular}{c}
			\includegraphics[scale=0.8]{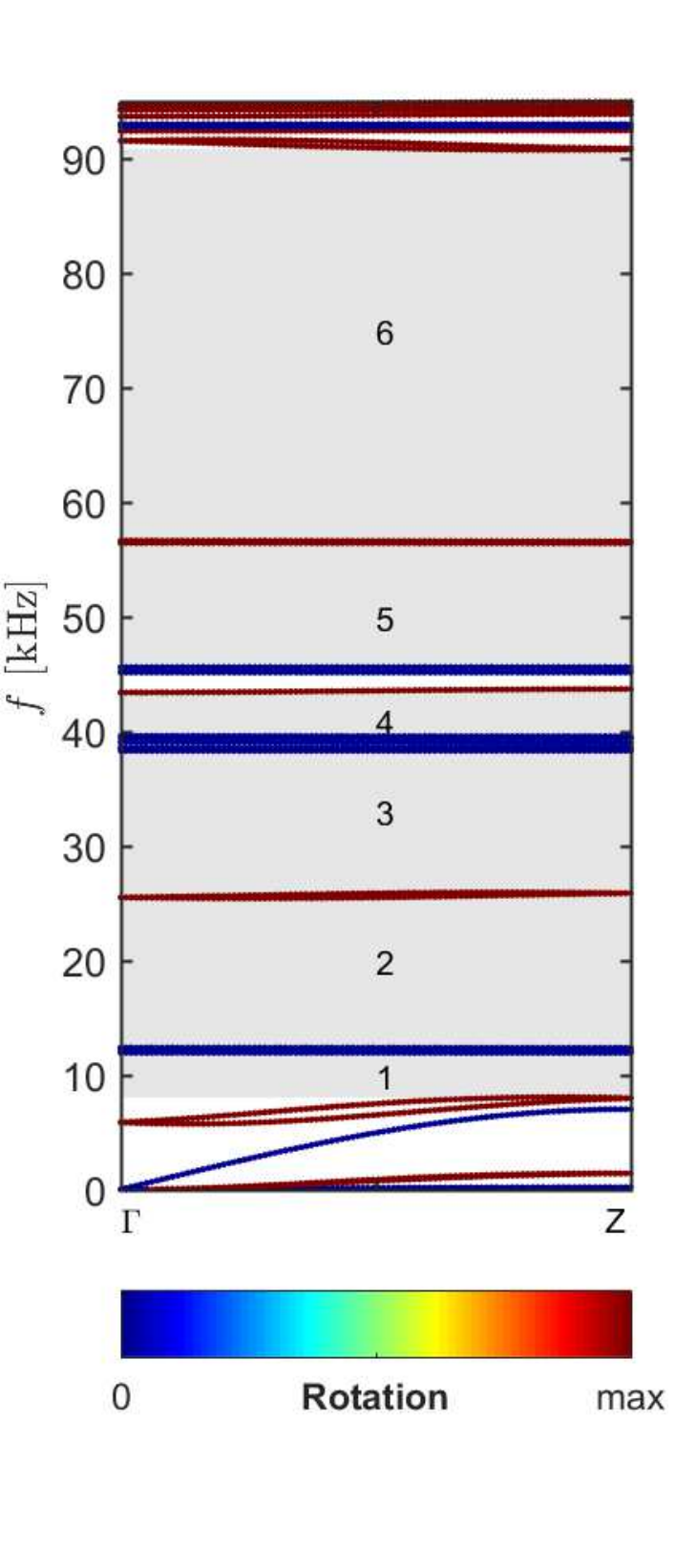}
		\end{tabular}
		& %[height=12.5cm,scale=0.65]
		\begin{tabular}{cc}
			\includegraphics[scale=0.12]{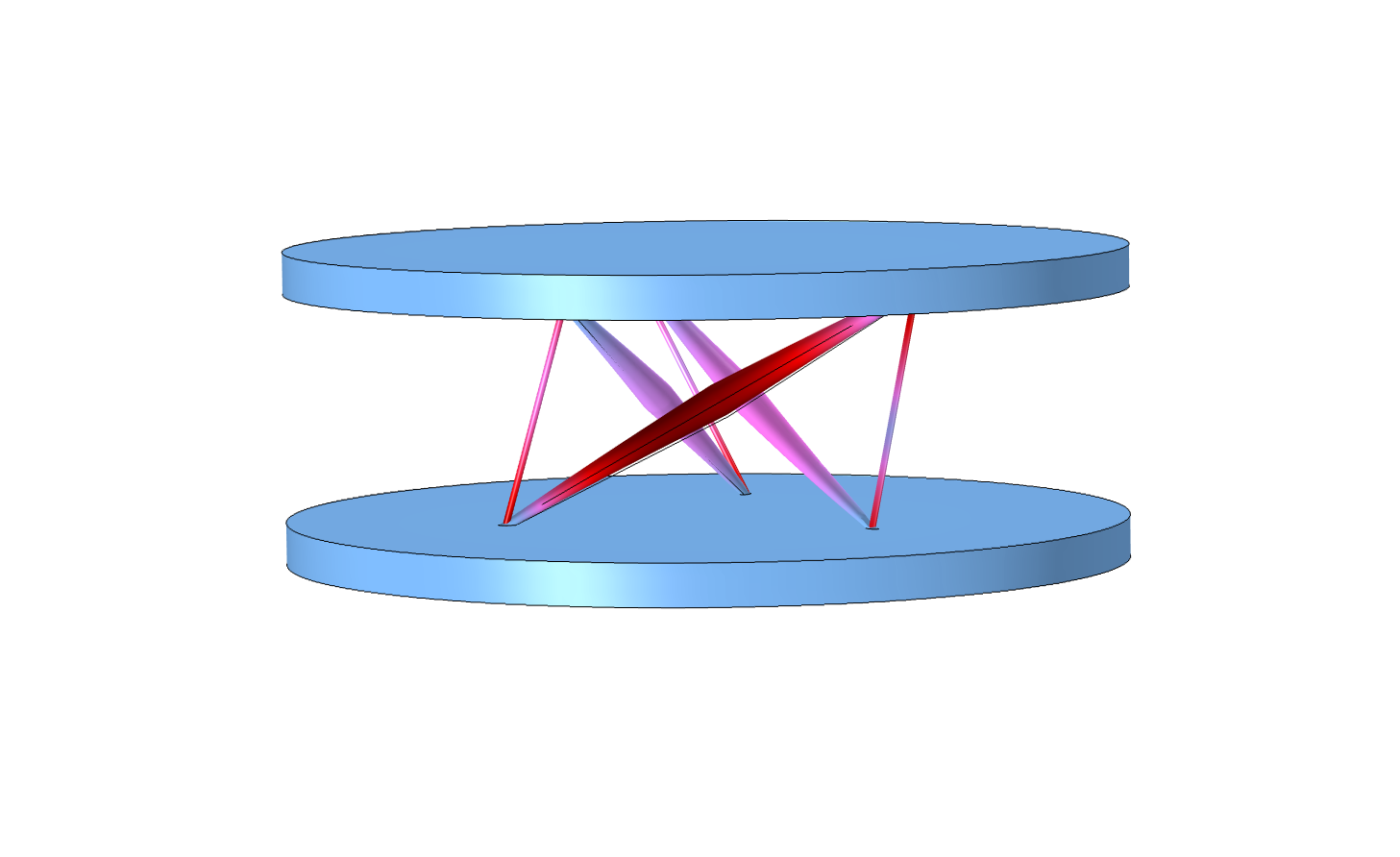}&
			\includegraphics[scale=0.12]{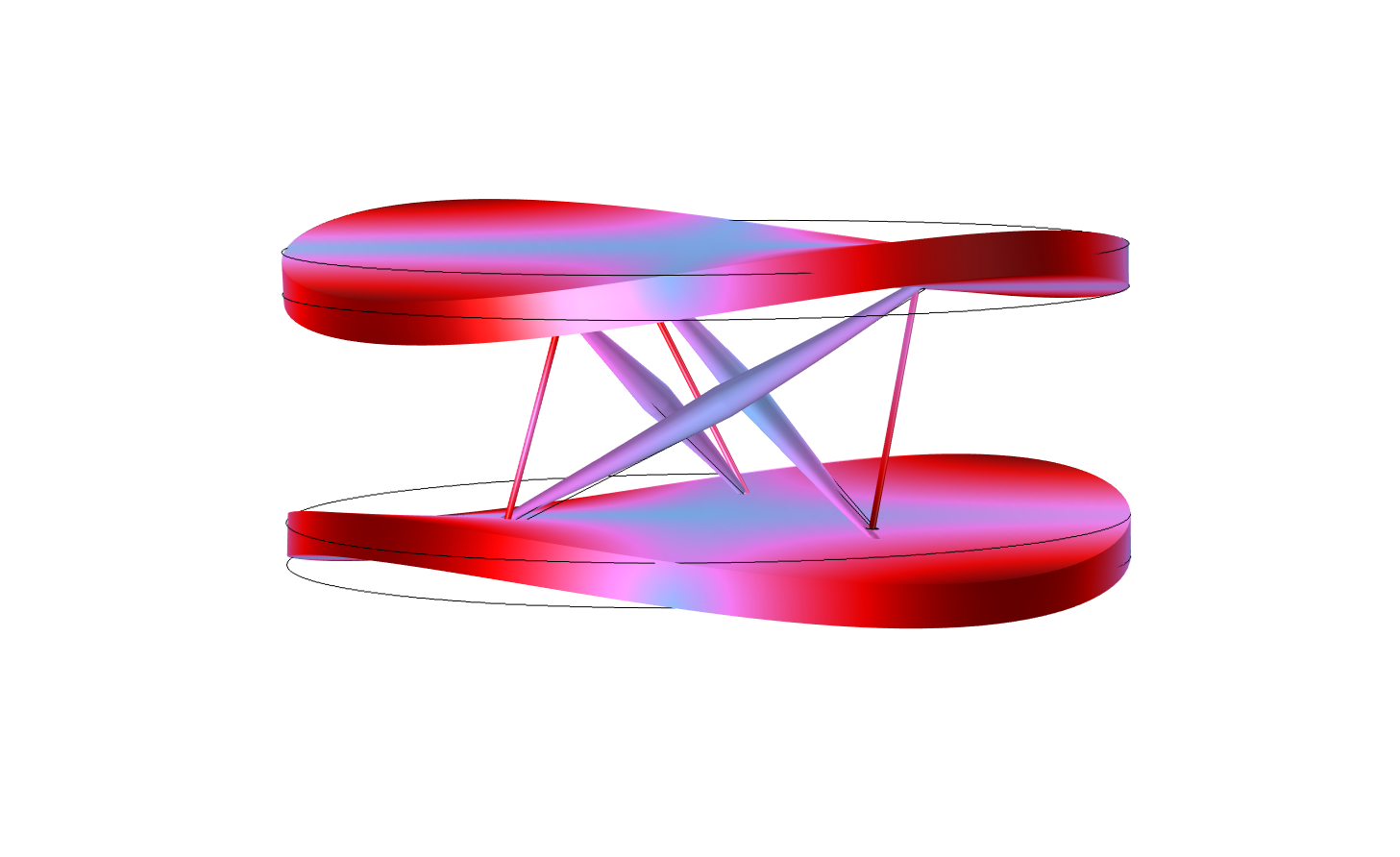}\\
			(b) 12.05 kHz& (c) 25.86 kHz\\
			\includegraphics[scale=0.08]{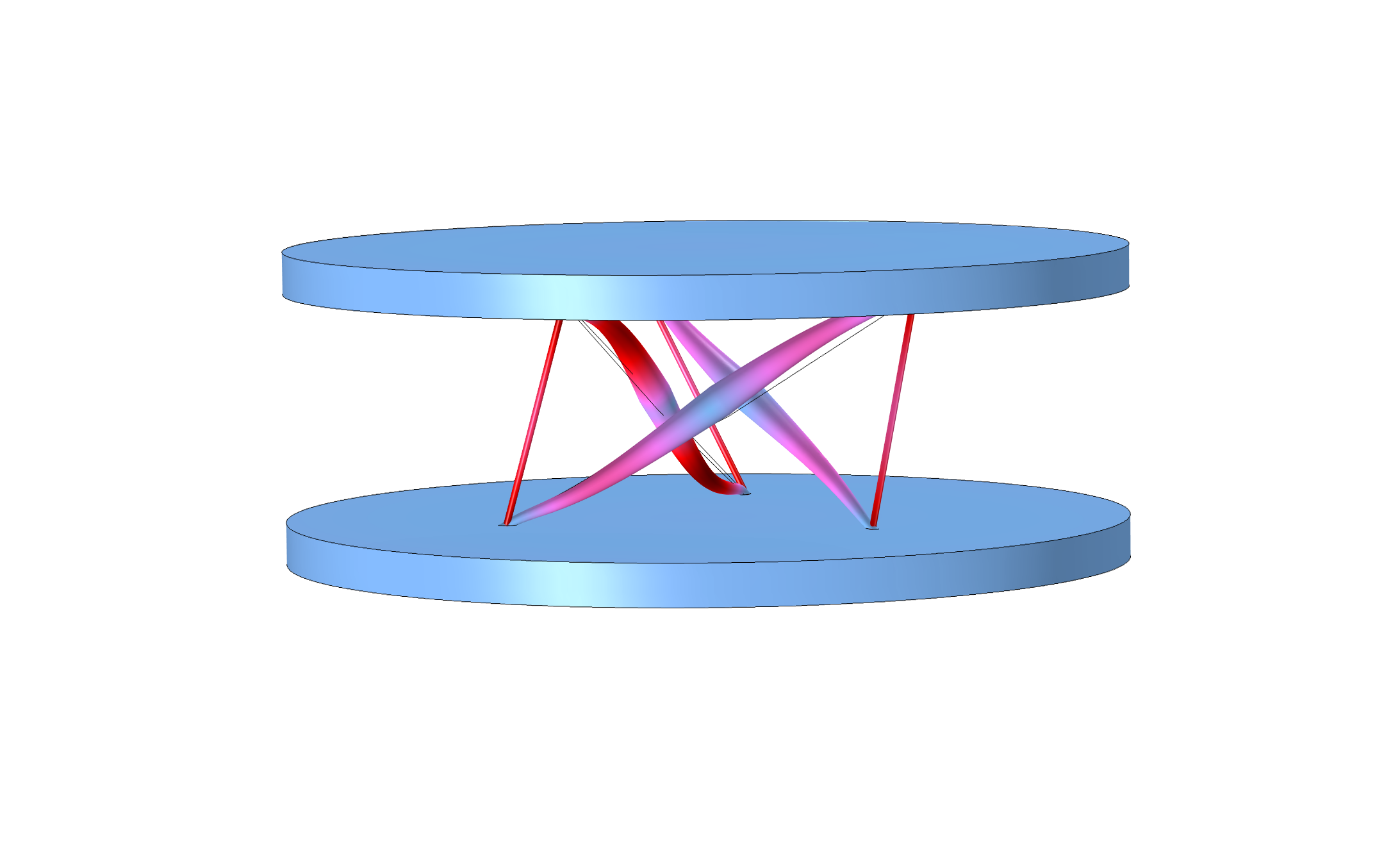}&
			\includegraphics[scale=0.08]{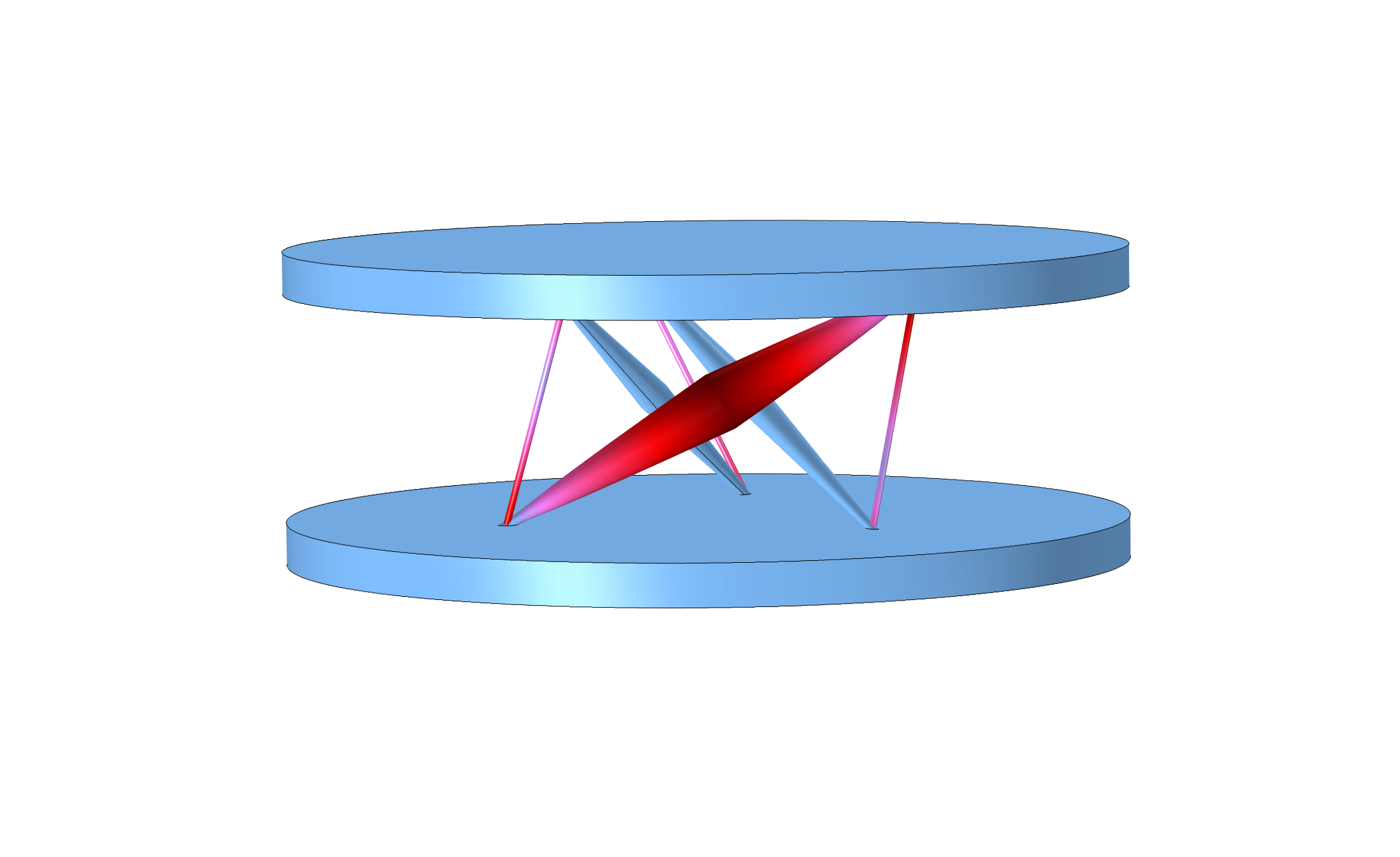}\\
			(d) 39.37 kHz& (e) 45.60 kHz\\
			\includegraphics[scale=0.08]{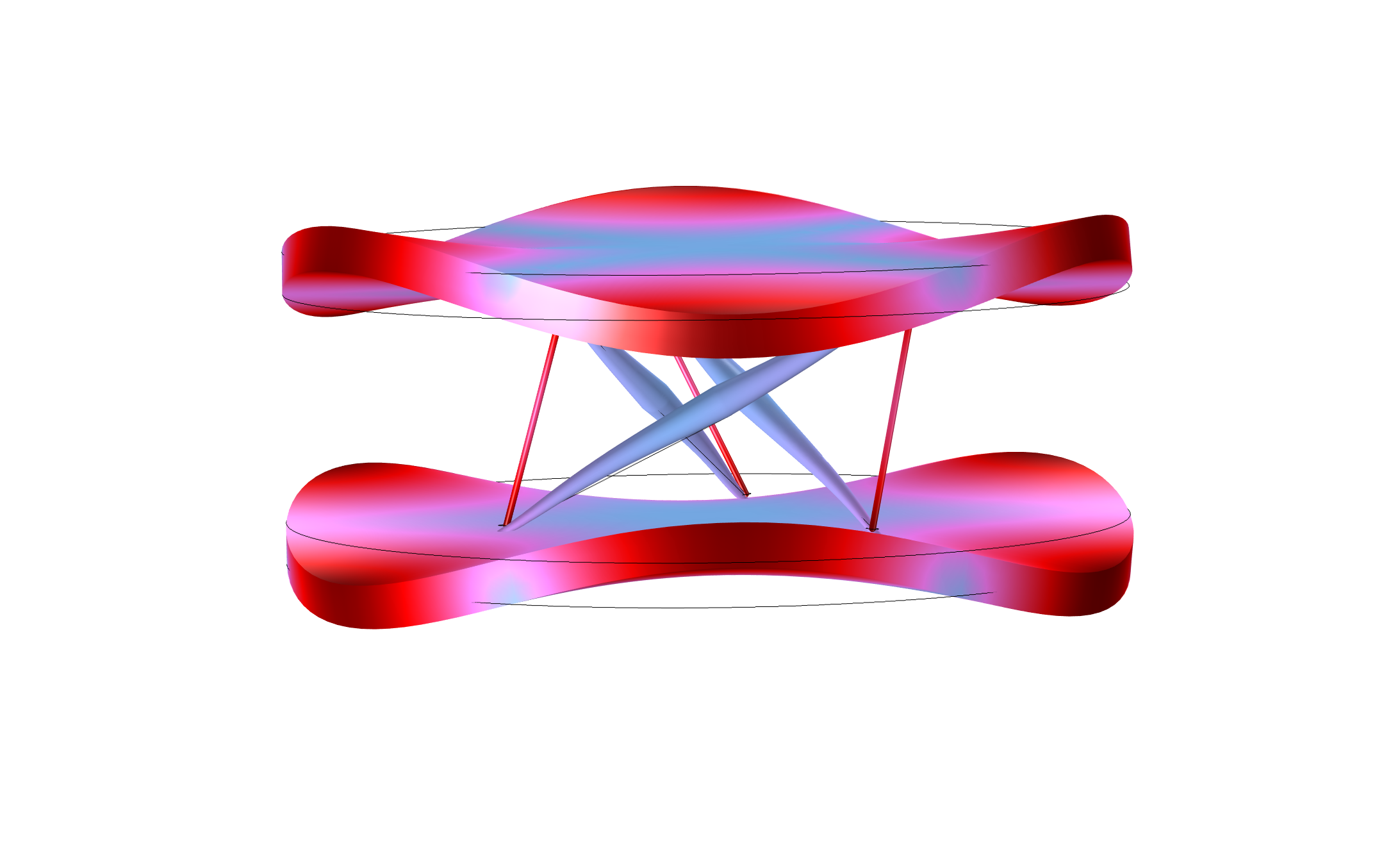}&
			\includegraphics[scale=0.08]{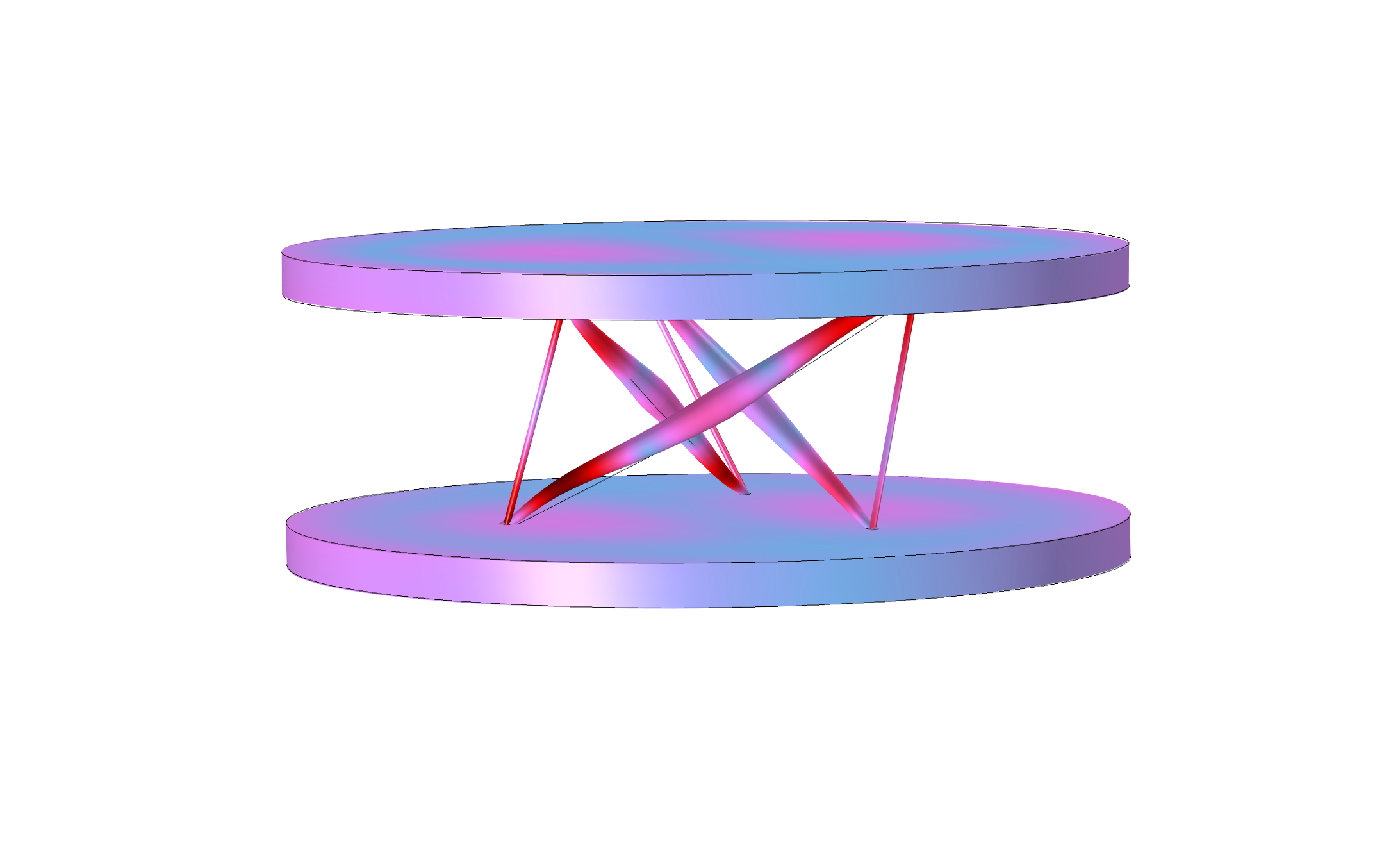}\\
			(f) 56.57 kHz& (g) 90.88 kHz\\
			\multicolumn{2}{l}{\qquad\qquad\qquad\includegraphics[scale=0.5]{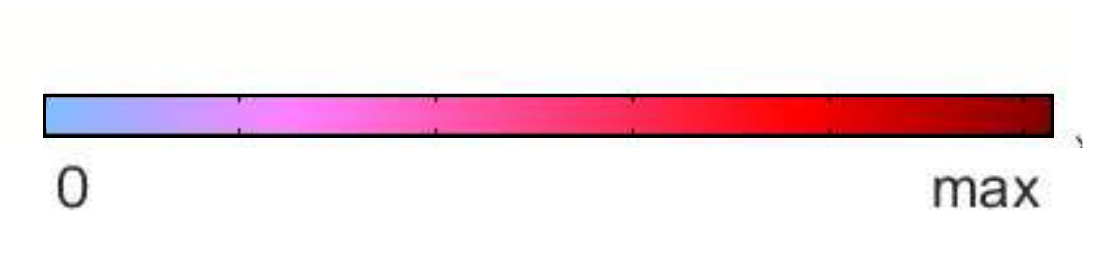}}
		\end{tabular}
	\end{tabular}
	\caption{{\bf Elastic model}: (a) 
		Dispersion relation for an accordion-like meta-chain with elastic discs of radius $R=10$~mm and thickness $t=2$~mm. Shaded regions numbered from 1 to 6 denote band gaps.
		Mode shapes for (almost) flat dispersion curves are evaluated at the point $Z$ of the Brillouin zone. They correspond to modes with vibrations localized in the bars (b,d,e,g) and in the discs (c,f).}
	\label{fig:ref_spectrum}
\end{figure}

Figure~\ref{fig:ref_spectrum}a presents the dispersion relation for a chain with the disc parameters $R$=10 mm, $t$=2~mm.
The color of the dispersion curves indicates the degree of bending rotations of the discs relative to the horizontal axes evaluated as:
%--------------------------------
\begin{equation}
\frac{\int\limits_A (\left|\omega_{x}\right|^2 + \left|\omega_{y}\right|^2) dA}{\int\limits_A(\left|\omega_{x}\right|^2 + \left|\omega_{y}\right|^2 + \left|\omega_{z}\right|^2) dA}=\frac{\int\limits_A (\left|\omega_{xk_z}\right|^2 + \left| \omega_{yk_z}\right|^2) dA}{\int\limits_A(\left|\omega_{xk_z}\right|^2 + \left|\omega_{yk_z}\right|^2 + \left| \omega_{zk_z}\right|^2) dA},
\end{equation}
%--------------------------------
where
%--------------------------------
\begin{eqnarray}
&&\left| \omega_{x}\right|^2 = \omega_x\bar{\omega}_x = \left(\frac{\partial u_z}{\partial y}-\frac{\partial u_y}{\partial z}\right)\left(\frac{\partial \bar{u}_z}{\partial y}-\frac{\partial \bar{u}_y}{\partial z}\right);\nonumber \\
&&\left|\omega_{y}\right|^2 = \omega_y\bar{\omega}_y = \left(\frac{\partial u_x}{\partial z}-\frac{\partial u_z}{\partial x}\right)\left(\frac{\partial \bar{u}_x}{\partial z}-\frac{\partial \bar{u}_z}{\partial x}\right);\nonumber\\
&&\left|\omega_{z}\right|^2 = \omega_z\bar{\omega}_z = \left(\frac{\partial u_y}{\partial x}-\frac{\partial u_x}{\partial y}\right)\left(\frac{\partial \bar{u}_y}{\partial x}-\frac{\partial \bar{u}_x}{\partial y}\right).
\end{eqnarray}
Here, $A$ is the volume of a disc, and the superimposed bar indicates complex conjugation. The subscript ${k_z}$ indicates that component $\omega_i$ (with $i$ denoting $x$, $y$ or $z$) is evaluated for each specified value of the wavenumber $k_z$.
In the colorbar of Fig.~\ref{fig:ref_spectrum}a, values range from 0 (blue) to `max' (red) indicating the change in mode polarization from translational to rotational.

%------------------------------------------------
%------------------------------------------------

Below the first band gap, the dispersion relation for the elastic model is identical to that of the rigid-elastic case (Fig.~\ref{fig:RigEl_spectrum}a). Hence, we  conclude that at these frequencies, the meta-chain dynamics is governed solely by the mass of a disk, and not its stiffness.
At higher frequencies, one observes multiple band gaps numbered from 1 to 6.
The upper bound of the first band gap is formed by a localized mode with displacements concentrated in the bars  (Fig.~\ref{fig:ref_spectrum}b). 
The other band gaps are also delimited by (almost) flat dispersion curves describing  localized modes. Some of these modes represent vibration modes of the inclined bars with (almost) motionless discs (Fig~\ref{fig:ref_spectrum}d,e,g), while the others describe higher-order vibrations of discs with bars at rest (Fig~\ref{fig:ref_spectrum}c,f). The localized character of motions and the flat shape of the mentioned dispersion curves indicate that the higher-frequency band gaps can be generated due to a locally resonant mechanism. The band-gap mechanisms is also discussed below.

%------------------------------------------------
\begin{figure}
	\includegraphics[scale=0.8]{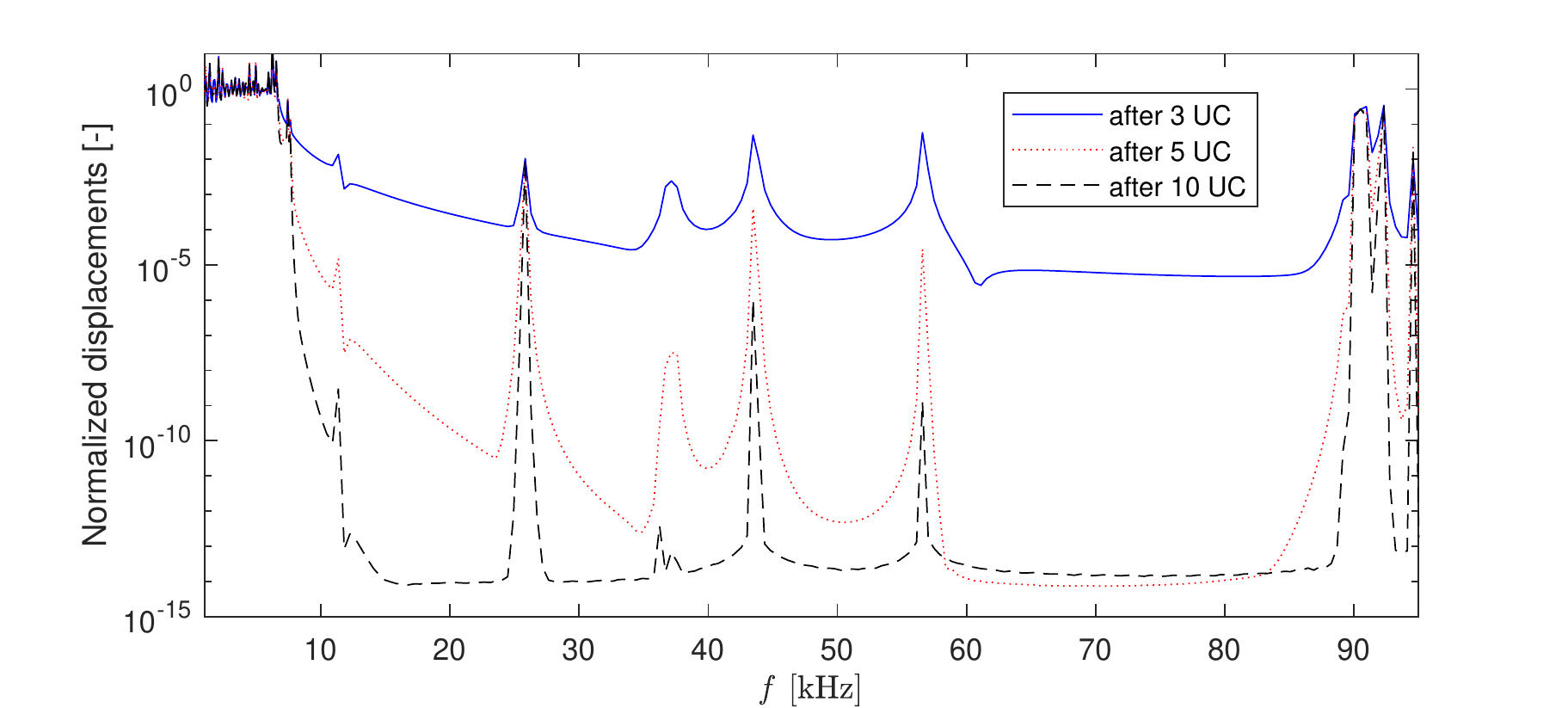}
	\caption{{\bf Elastic model:} Normalized transmitted displacements $\sqrt{u_x^2+u_y^2+u_z^2}/u_{z0}$ versus frequency $f$ for an accordion-like meta-chain composed of 30 unit cells (disc dimensions are $R=10$~mm, $t=2$~mm). The curves indicate displacements averaged on three adjacent discs located at distances of 3,5, or 10 unit cells from the loaded end of the chain.}
	\label{fig_trans_elast}
\end{figure}
%------------------------------------------------

For the six band gaps in Fig~\ref{fig:ref_spectrum}a, the ratio between the gap width and the mid-gap frequency 
is 38.2\%, 69.3\%. 38.8\%, 9.1\%, 21.2\%, and 46.3\%, respectively. However, the
localized character of the modes suggests that one can expect inefficient excitation of some of the modes under certain loading conditions, and, as a result, extension of the band-gap width.
To verify this assumption and further understand the band-gap origins, we perform  transmission analysis
by using a similar model and excitation loading as in the corresponding rigid-elastic case (Section~\ref{sec:rigid-elastic}).
Figure~\ref{fig_trans_elast} shows the normalized magnitude of the transmitted total displacement averaged on three adjacent discs for the uniform axial excitation. At the band-gap frequencies, waves attenuate significantly while propagating through 5 unit cells. The attenuation is stronger compared to the rigid-elastic case, since the wave velocity in an elastic material is lower.
The modes characterized by localized motions in bars (flat blue lines in Fig.~\ref{fig:ref_spectrum}a with mode shapes in  Figs.~\ref{fig:ref_spectrum}b,d,e,g) are strongly attenuated. This means that the applied displacement cannot efficiently excite them. In general,
it is very difficult to induce vibrations in isolated bars, while the discs are motionless. Hence, we conclude that the band gaps $\#$1 and $\#$2 as well $\#$3 and $\#$4 are effectively merged into two wide band gaps with BG\% of 103.6\% and 50.6\%, respectively.

Therefore, the transmission analysis results reveal that the accordion-like meta-chains are capable of strongly attenuating elastic waves in wide frequency ranges. Note that in contrast to locally resonant metamaterials with coated inclusions or pillars, wave attenuation within the band gaps is uniform,  not similar to asymmetric Fano-like profiles~\cite{liu2000}. This suggests that the second and higher band gaps are generated due to a coupling between the Bragg scattering, occurring within the first band gap, and localized modes of different metamaterial constituents. The coupled band-gap mechanism enables to overcome an inherently narrow-frequency attenuation limit typical for the local resonance mechanism. Moreover, in combination with rich design possibilities of tensegrity structures it provide a powerful means to tune band gap in wide frequency ranges by altering the structural geometry, as shown in Section~\ref{sec:param_studies}.

The designed meta-chains are  lightweight  with the material filling fraction of $27$\% and a broad (merged) low-frequency band gap of width exceeding 100\%.
As compared to other configurations shown in Fig.~\ref{fig_BGsize}, they have the lowest volume fraction and exhibit band gaps at the same or lower frequencies. In addition, the first band gap is accompanied by several wide higher-frequency gaps that is not the case for most optimized designs~\cite{bh11, dac16, laude2017}.

%% BG sizes
% 1st:  8100 - 11933 Hz  BG%: 38.2%
% 2nd: 12375 - 25493 Hz  BG%: 69.3%
% 3rd: 25861 - 38295 Hz  BG%: 38.8%
% 4th: 39595 - 43388 Hz  BG%:  9.1%
% 5th: 45598 - 56424 Hz  BG%: 21.2%
% 6th: 56634 - 90782 Hz  BG%: 46.3%

%% BG sizes, p=0.001
% 1st:  7620 - 11017 Hz  BG%: 36.5%
% 2nd: 11546 - 25483 Hz  BG%: 75.3%
% 3rd: 25818 - 34979 Hz  BG%: 30.1%
% 4th: 38891 - 43371 Hz  BG%: 10.9%
% 5th: 43520 - 56421 Hz  BG%: 25.8%
% 6th: 56619 - 88587 Hz  BG%: 44.0%

%% BG sizes, p=0.005

\begin{figure}
	\centering
	\includegraphics[scale=0.8]{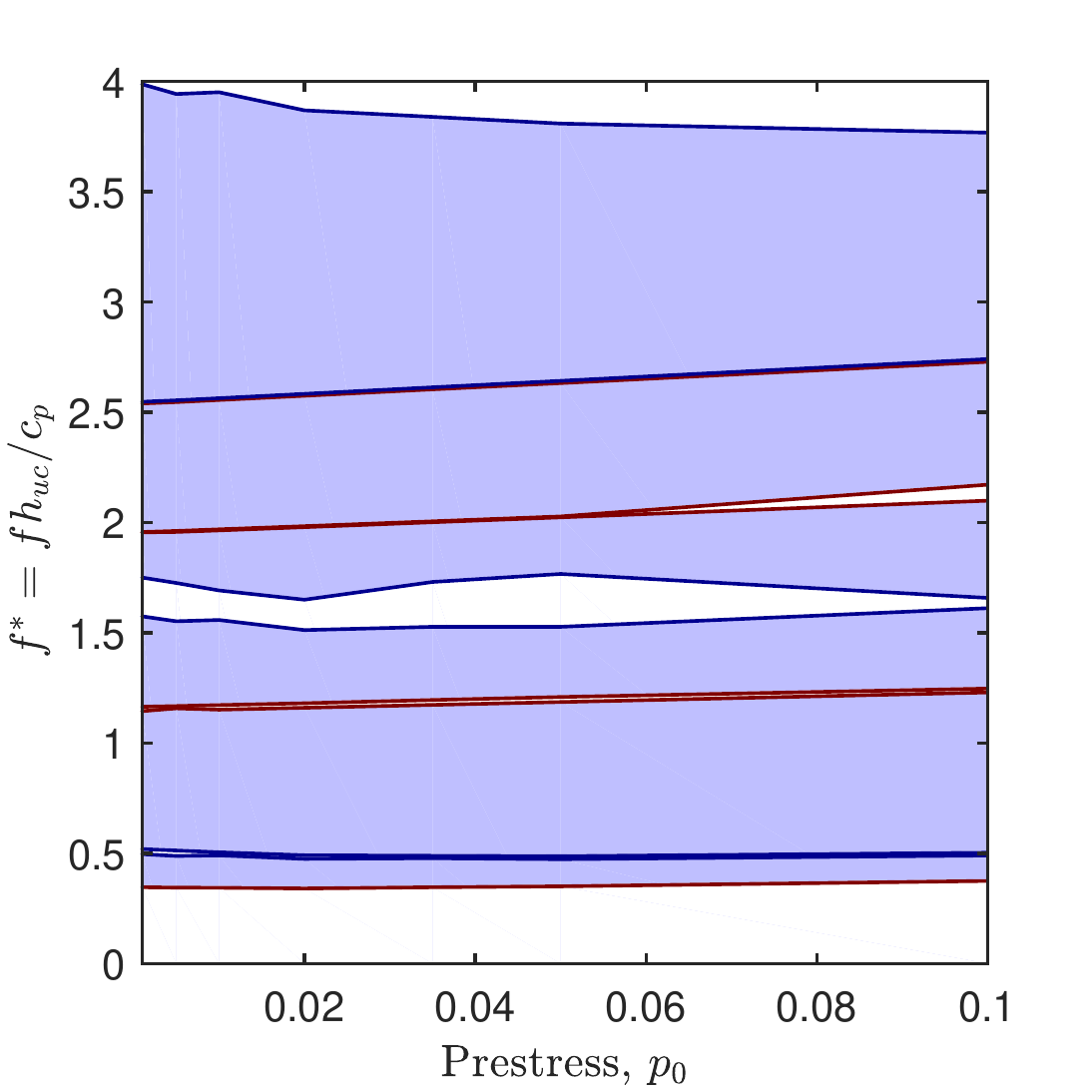}
	\caption{The band-gap frequencies {\em vs.} the level of internal prestress $p_0$ in the cross-strings for an  accordion-like meta-chain.}
	\label{fig: pre-stress data}
\end{figure}

The analysis of the mode shapes in Fig.~\ref{fig:ref_spectrum}~b-g indicates that the cross-strings in tensegrity lattices undergo large tensile deformations, while waves propagate along a meta-chain. However, their role on the wave dispersion is negligible due their lower material stiffness.
If prestress is applied, the height of a tensegrity lattice increases. This shifts the edges of the pass bands, as shown in Fig.~\ref{fig: pre-stress data}.  Here we normalize the frequency as $f^* = fh_{uc}/c_p$, where $c_p$ is the effective phase velocity evaluated from the dispersion relation as described in Section~\ref{sec:rigid-elastic}. 
Possible approaches to introduce internal prestress include either the application of external mechanical forces {\em in situ} or 
the use of micro-stereolitography setups for manufacturing the strings. 
The latter use swelling materials that contract, once dried, and thus, create internal prestress  (see Ref.~\cite{aa2018} and the references therein). The exploitation of materials with different thermal-expansion coefficients opens a way to control the level of prestress by varying the ambient temperature.

%------------------------------------------------
%------------------------------------------------

\section{Parametric studies of the wave dispersion for an elastic meta-chain} \label{sec:param_studies}

The choice of geometric parameters for the 
metamaterial unit cells is in part governed by previously analyzed structures in the literature~\cite{fra14,aa2018}, but is also in part arbitrary.
To demonstrate that the found band gaps are preserved for a wide range of the parameters and can be tuned to desired frequencies by varying the geometric and structural parameters, we perform
a series of numerical studies.
For this purpose, we consider the elastic meta-chain model 
and focus our attention on a low-frequency range, including the first three band gaps, as mostly promising for applications.

We start from variations of the disc mass by changing the disc radius while preserving its thickness and then vary the disc material density maintaining the same disc geometry. 
After this, we estimate the influence of the end diameter of the bars on band-gap frequencies. All these simulations are performed for zero initial prestress. 

\subsection{Disc radius variations}

\begin{figure}[h]
	\begin{tabular}{cc}
		\subfloat[]{\includegraphics[scale=0.5]{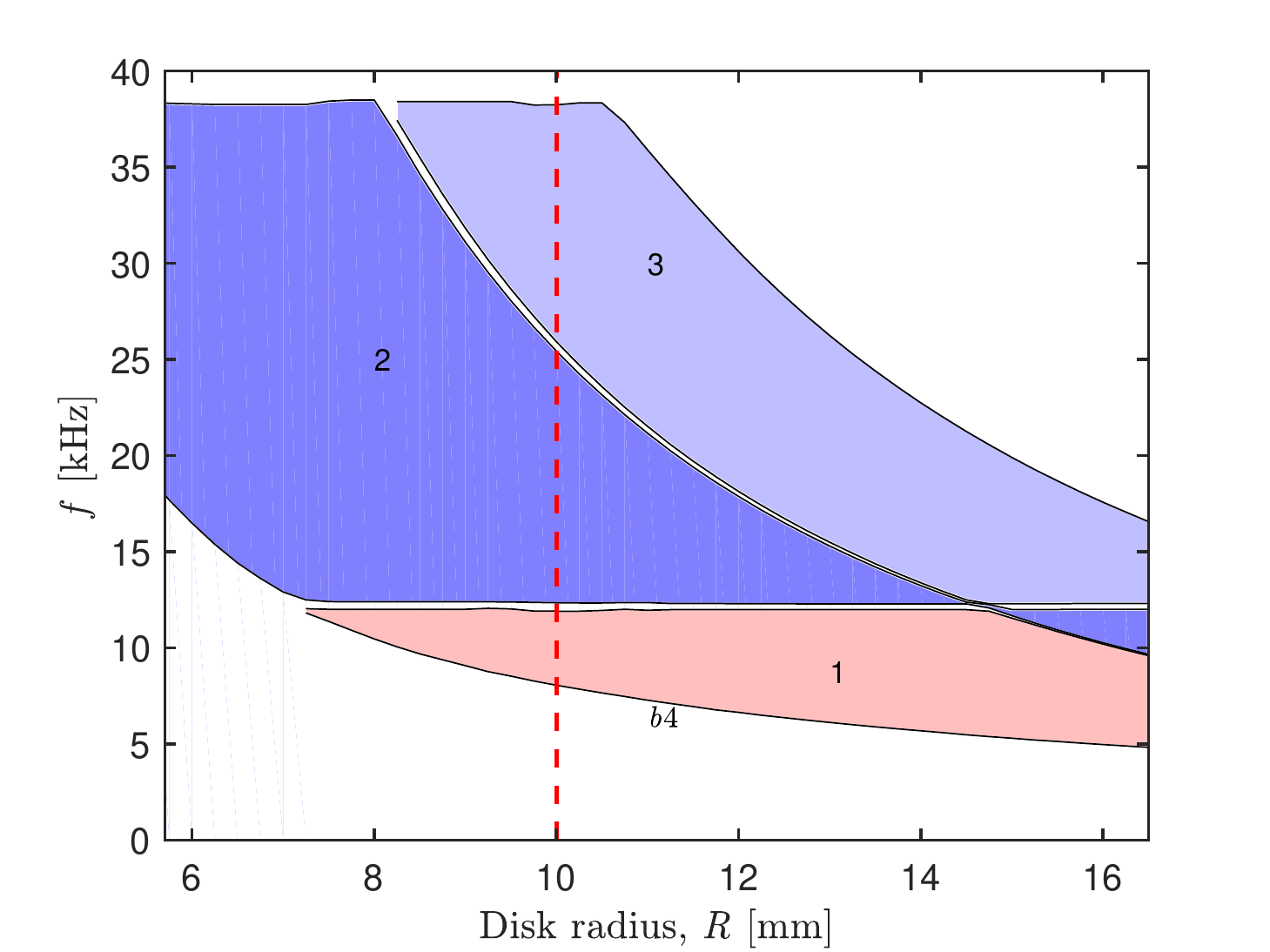}}&
		\subfloat[]{\includegraphics[scale=0.5]{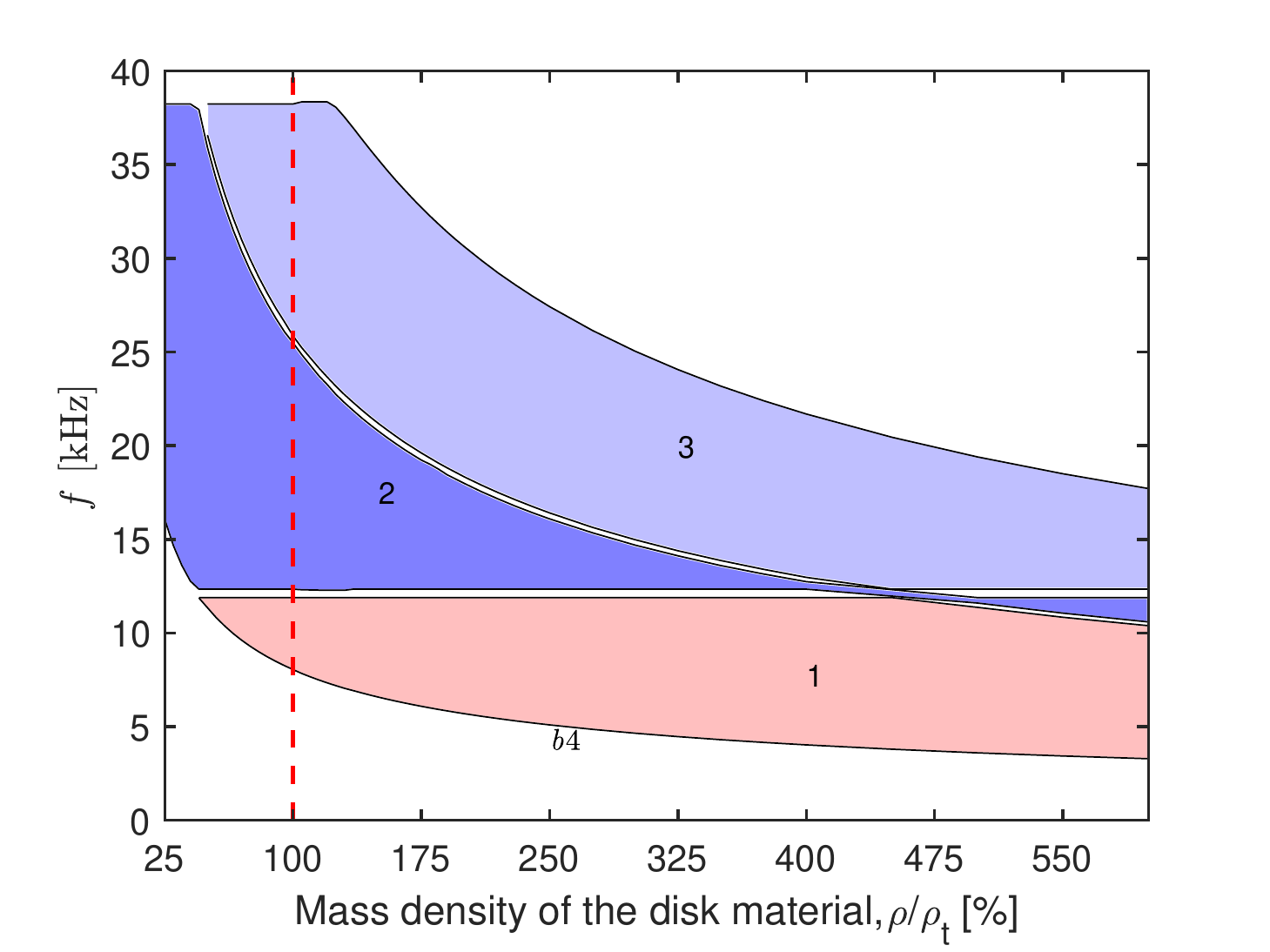}}
	\end{tabular}
	\caption{Frequencies of the three lowest band gaps (numbered from 1 to 3) versus the radius (a) or
		the mass density  (b) of a disc.  The red dashed line indicates a reference configuration with the dispersion relation shown in Fig.~\ref{fig:ref_spectrum}a. The notation $b4$ refers to the fourth rotational mode that is the lower bound of the first band gap.}
	\label{fig:par_st_R_rho}
\end{figure}

Figure~\ref{fig:par_st_R_rho}a shows the band-gap frequencies (shaded regions) versus the disc radius $R$. The corresponding dispersion relations can be found in Video 1. The red dashed line indicates the reference case (Fig.~\ref{fig:ref_spectrum}a). 
For the minimum possible radius of the disc $R=5.5$~mm, governed by the lateral dimensions of a tensegrity prism, there is a single wide band gap. The unit-cell effective mass density and the material volume fraction are $\rho_{eff}=1250$~kg/m$^3$ and $28$\%, respectively. As $R$ increases and the effective density decreases, the band gap is shifted to lower frequencies and splits into smaller ones due to the appearance of localized and dispersional modes. These band gaps are located close to each other and can be merged if some of the modes are not efficiently excited, similarly to the case described in Section~\ref{sec: elastic case}. For the largest analyzed value of $R=16.5$~mm and $\rho_{eff}=1206$ kg/m$^3$.

%----------------------------------
%---------------------------

Different masses of the terminal discs enable generation of multiple band gaps.
As can be expected, the increase of the disc radius (and thus, its mass) shifts the band gaps to lower frequencies. However, it is accompanied by excitation of localized and bending modes, which decrease the gap width (see Video 1 for more details). In this case, one can extend attenuation frequency ranges by exploiting a cascade design for the disc masses~\cite{ktd15}.

\subsection{Disc mass effects}
The variation of the material mass density is equivalent to the analysis of non-homogeneous structures, in which 
the bars in a tensegrity lattice and the terminal discs are made of different materials.
To facilitate the comparison with the homogeneous case, we maintain the same disc's geometry, i.e., $R=10$~mm and $t=2$~mm.
This allows us to maintain the same effective stiffness of the structure and analyze only the mass variation.

Figure~\ref{fig:par_st_R_rho}b shows the band-gap width {\em vs.} the percent ratio between  mass density $\rho$ and mass density of titanium $\rho_t$. The dispersion relations can be found in Video 2. The red dashed line refers to the reference case of a homogeneous metamaterial (Fig.~\ref{fig:ref_spectrum}a). The 600\% increase of the disc mass shifts the first band gap to about twice lower frequencies.
Note that a general structure of band gaps in Figure~\ref{fig:par_st_R_rho}b is very similar to that in Figure~\ref{fig:par_st_R_rho}a.
Therefore, the mass of a disc and not its geometry play a crucial role in the generation of low-frequency band gaps. This conclusion can be expected from the fact that modes occurring at the band-gap bounds either imply motions of the discs as rigid bodies or involve localized vibration the bars. However, at higher frequencies, the disc dynamics  obviously plays an important role.

\subsection{Variation of the bar geometry}
%---------------------------
\begin{figure}
	\begin{tabular}{cc}
		\begin{tabular}{c}
			\subfloat[]{\includegraphics[scale=0.55]{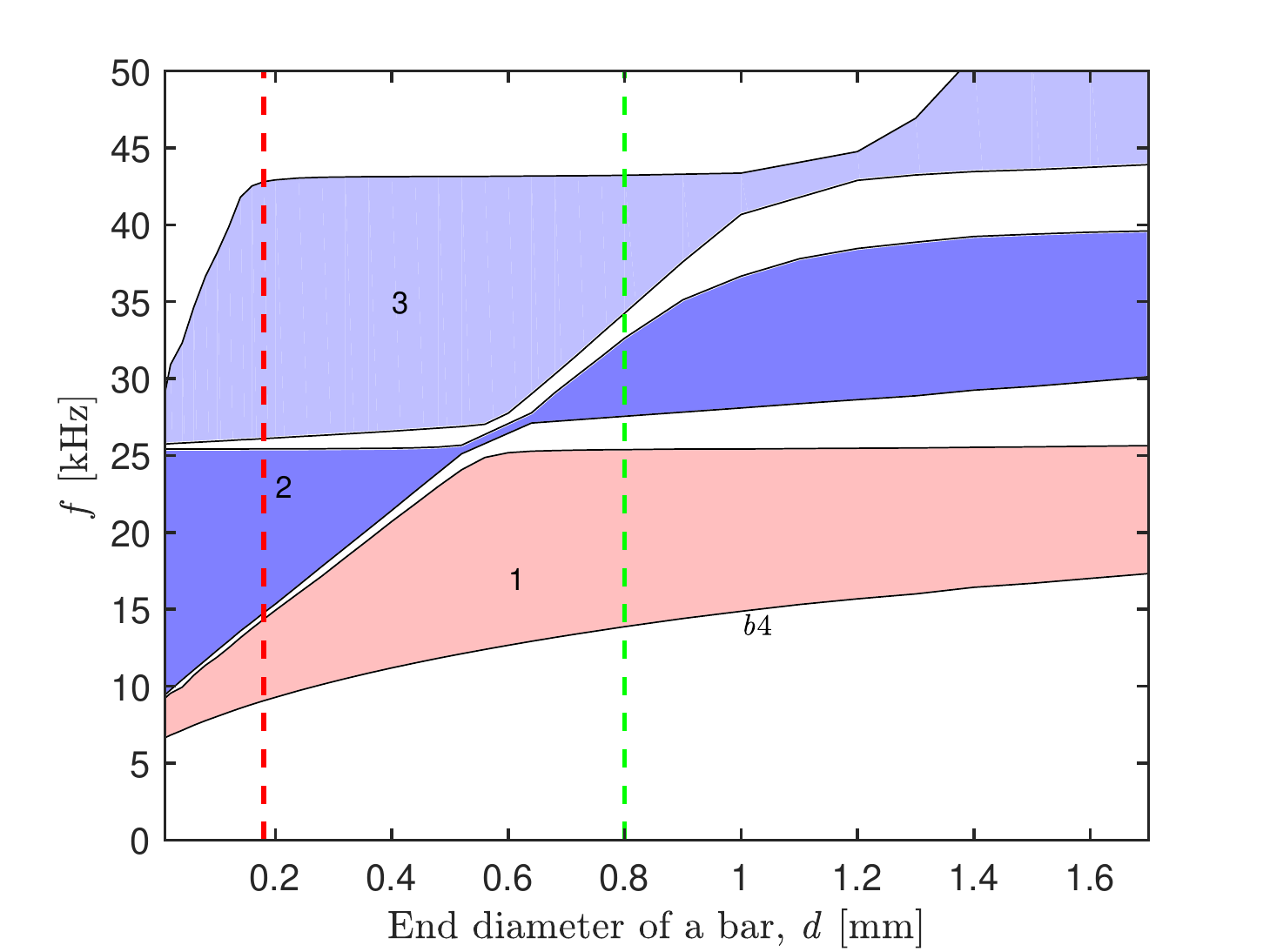}}
		\end{tabular}
		&
		\begin{tabular}{c}
			\subfloat[]{\includegraphics[scale=0.085]{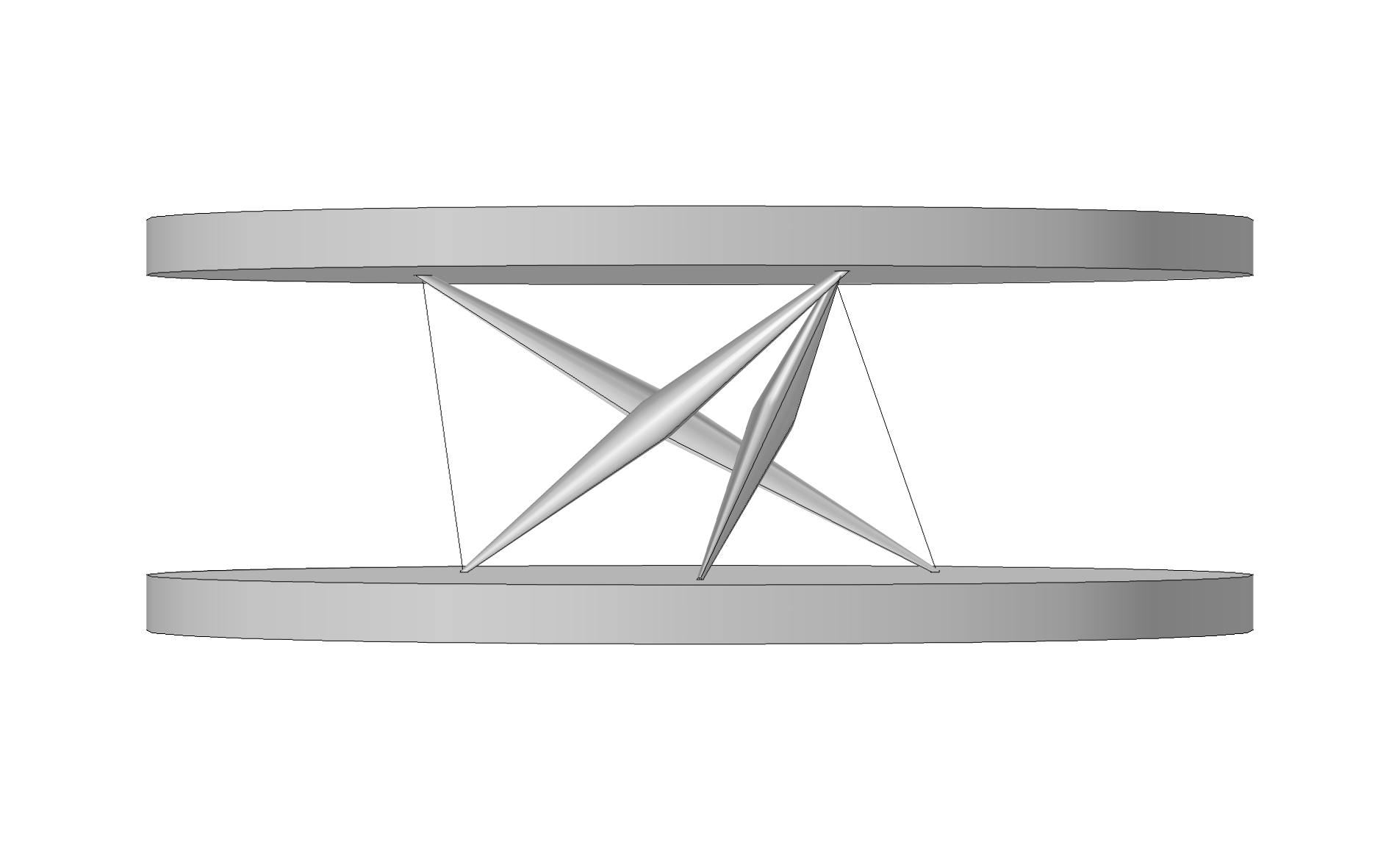}}\\
			\subfloat[]{\includegraphics[scale=0.085]{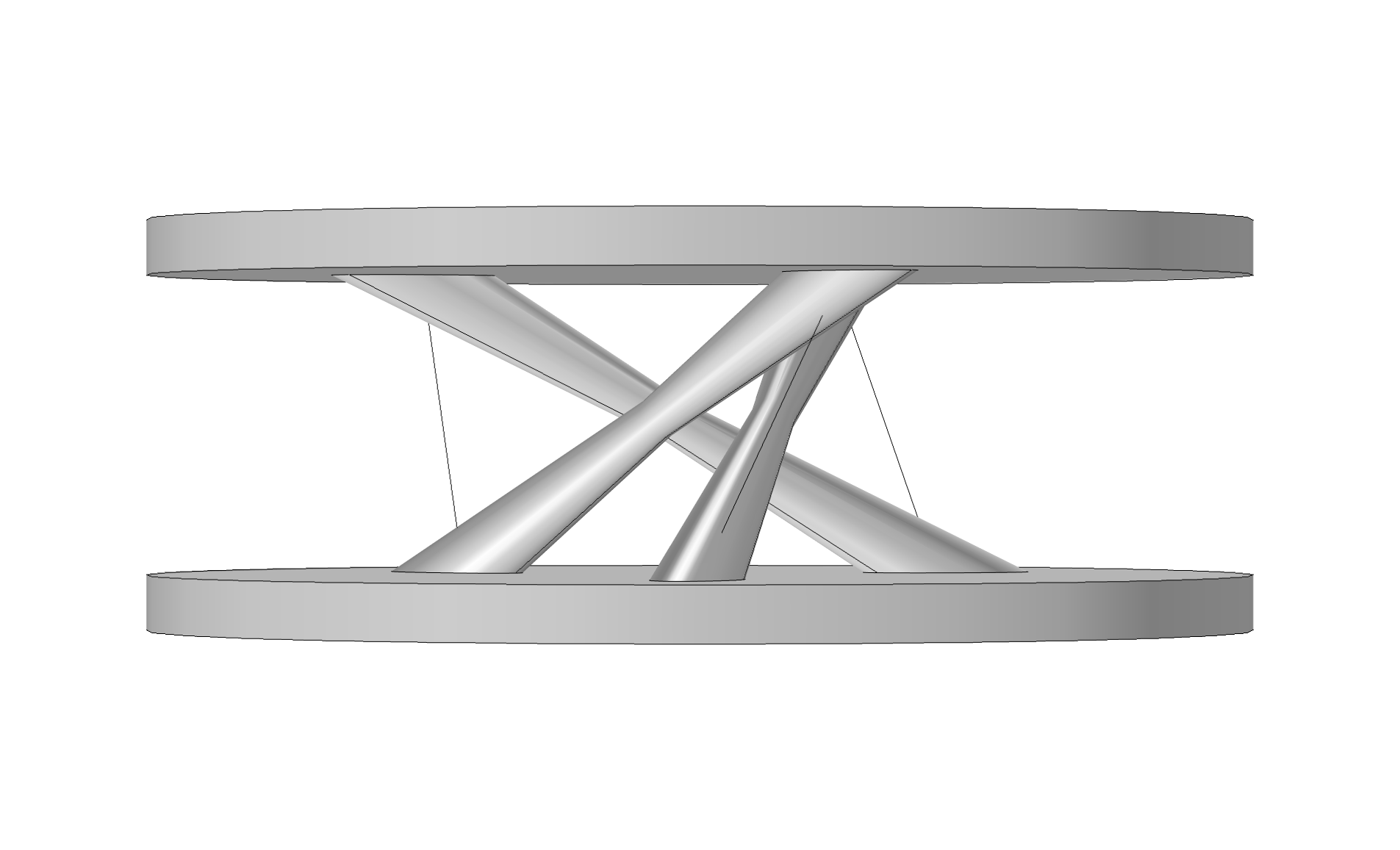}}
		\end{tabular}
	\end{tabular}
	\caption{(a) Frequencies of the three lowest band gaps (numbered from 1 to 3) versus the end diameter of a bar in the tensegrity prism. The green and red dashed lines indicate the geometry with bars of a uniform cross section ($d=D$) and a reference configuration (Fig.~\ref{fig:ref_spectrum}a). The metamaterial unit cells with the end diameters $d=0.1$~mm (b) and $d=1.7$~mm (c).}
	\label{fig:param_st_d}	
\end{figure}
%---------------------------

As shown in Section~\ref{sec:rigid-elastic}, the unit-cell stiffness is governed by the tangential stiffness of inclined bars, and thus, is strongly dependent on bar geometry of a bar. To study this dependence, we vary the end diameter $d$ of a bar, while the central diameter $D$ is fixed, and evaluate dispersion relations (see Video 3).  

Figure~\ref{fig:param_st_d}a
shows the frequencies of the three band gaps for various values of $d$.
Green and red dashed lines indicate the meta-structure with bars of uniform cross section ($d=D$) and the reference case (Fig.~\ref{fig:ref_spectrum}a).
The unit cell geometries with the smallest and largest analyzed end diameters are shown in Figure~\ref{fig:param_st_d}b and c. 
In general, the three low-frequency band gaps are generated for all the geometries.
As expected, larger values of $d$ correspond to a stiffer unit-cell response, and as a result, the band gaps are located at higher frequencies.
For $d\ge 0.52$~mm, the modes characterized by intense vibrations in the bars are non-localized (Fig.~\ref{fig:ref_spectrum}b,d,e), since the discs have non-zero deformations due to a strong coupling with the bars. The larger the value of $d$, the larger is the distance between adjacent band gaps. Therefore, one can anticipate that the band gaps can be effectively
merged only for tapered bars, when the end diameter $d$ of a bar is smaller than its central diameter $D$.

\section{Three-dimensional accordion-like metamaterials}\label{sec:3Dmodel}
\begin{figure}[h]
	\begin{tabular}{c}
		\begin{tabular}{cc}
			\subfloat[]{\includegraphics[scale=0.15]{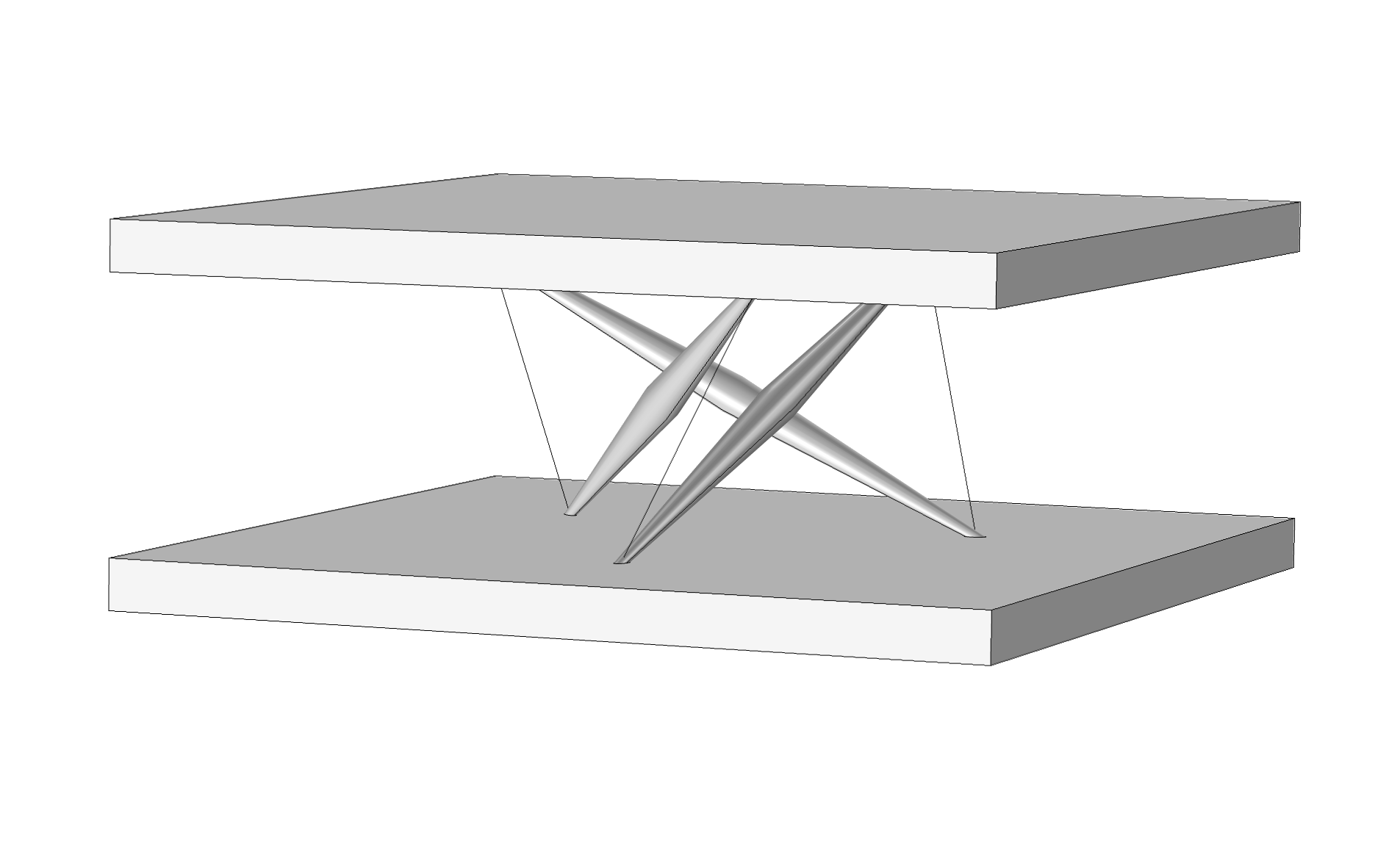}}&
			\subfloat[]{\includegraphics[scale=0.2]{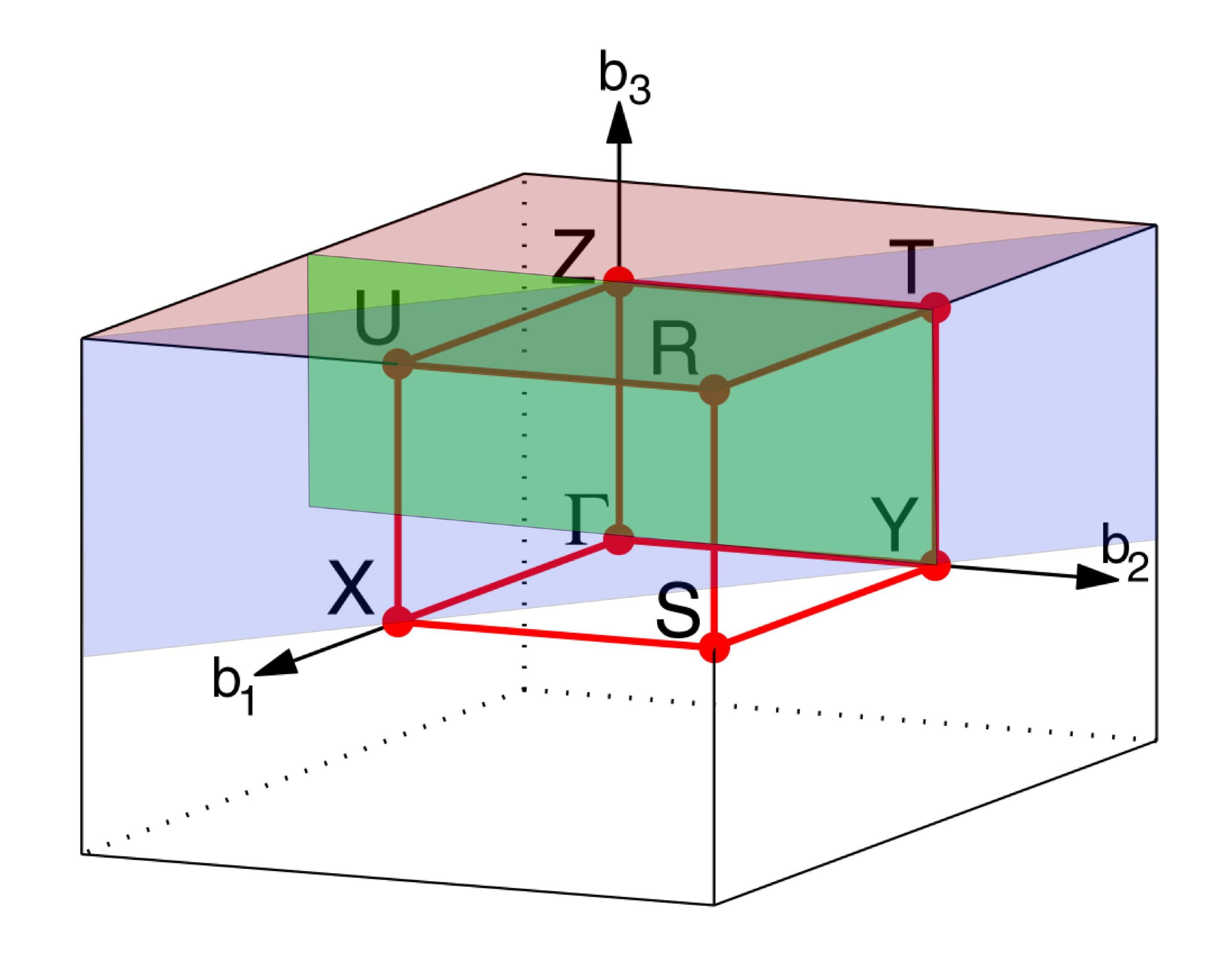}}
		\end{tabular}
		\\
		\begin{tabular}{cc}
			\begin{tabular}{c}
				\subfloat[]{\includegraphics[scale=0.8]{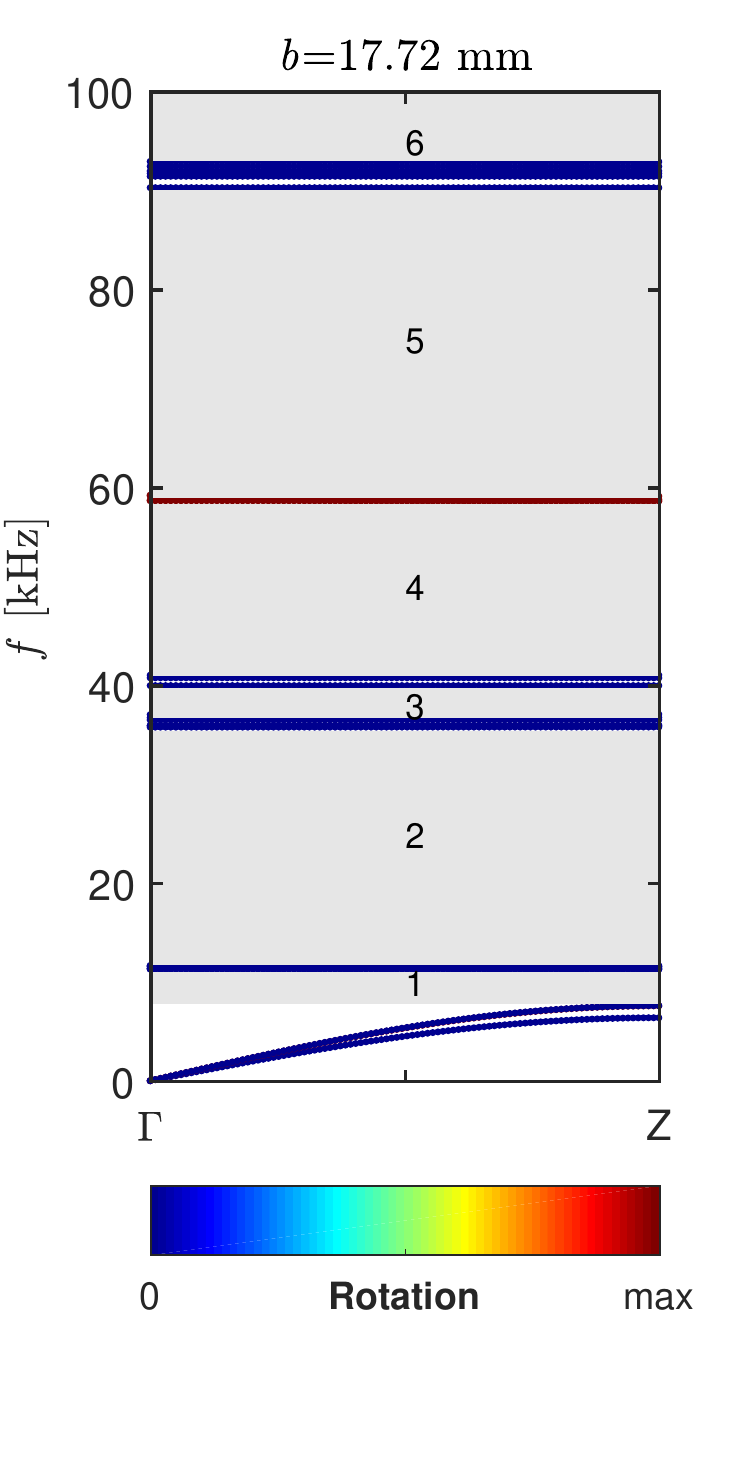}}
			\end{tabular}
			&
			\begin{tabular}{c}
				\includegraphics[scale=0.09]{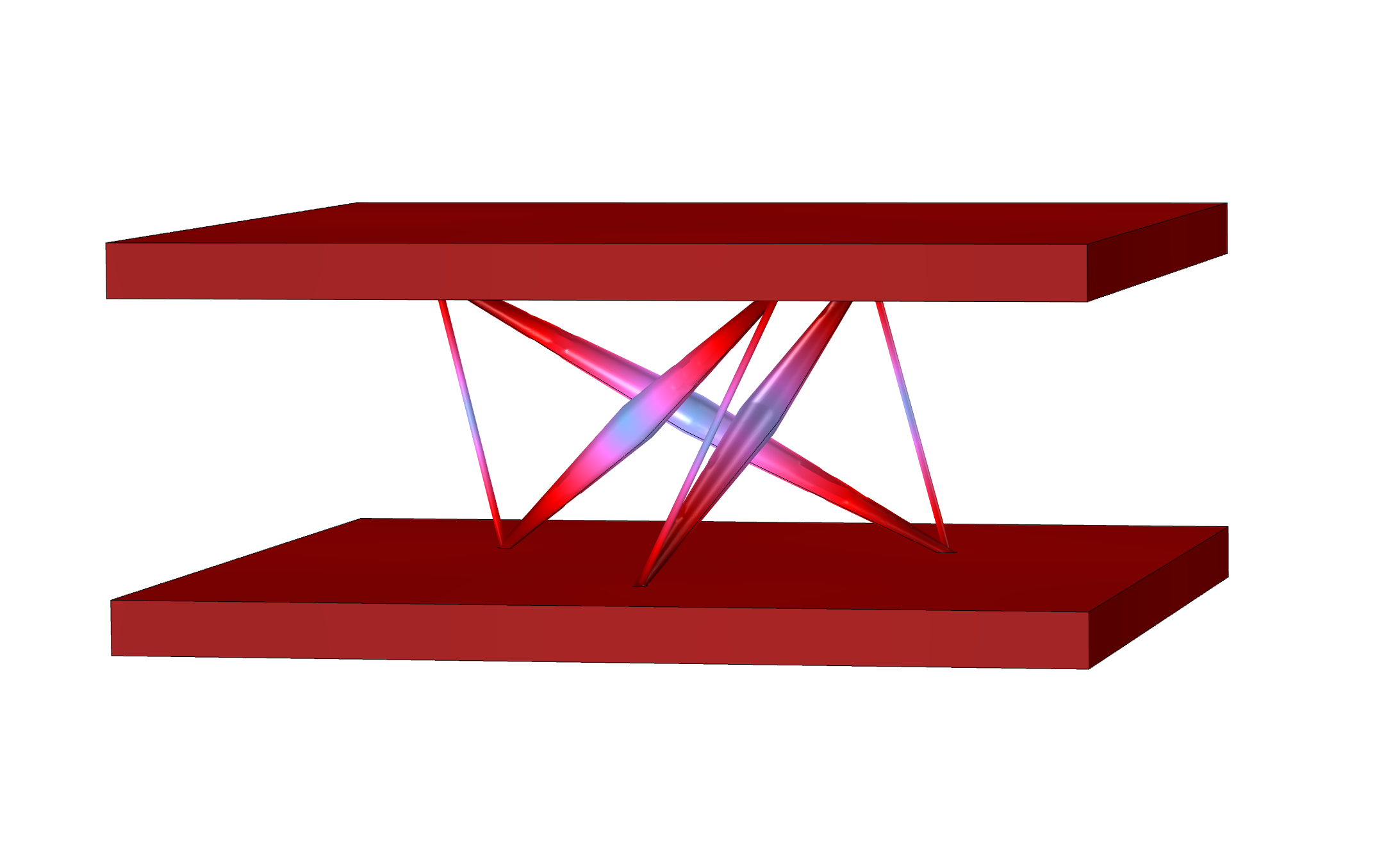}\\
				(d) $f=6.37$~kHz\\
				\includegraphics[scale=0.09]{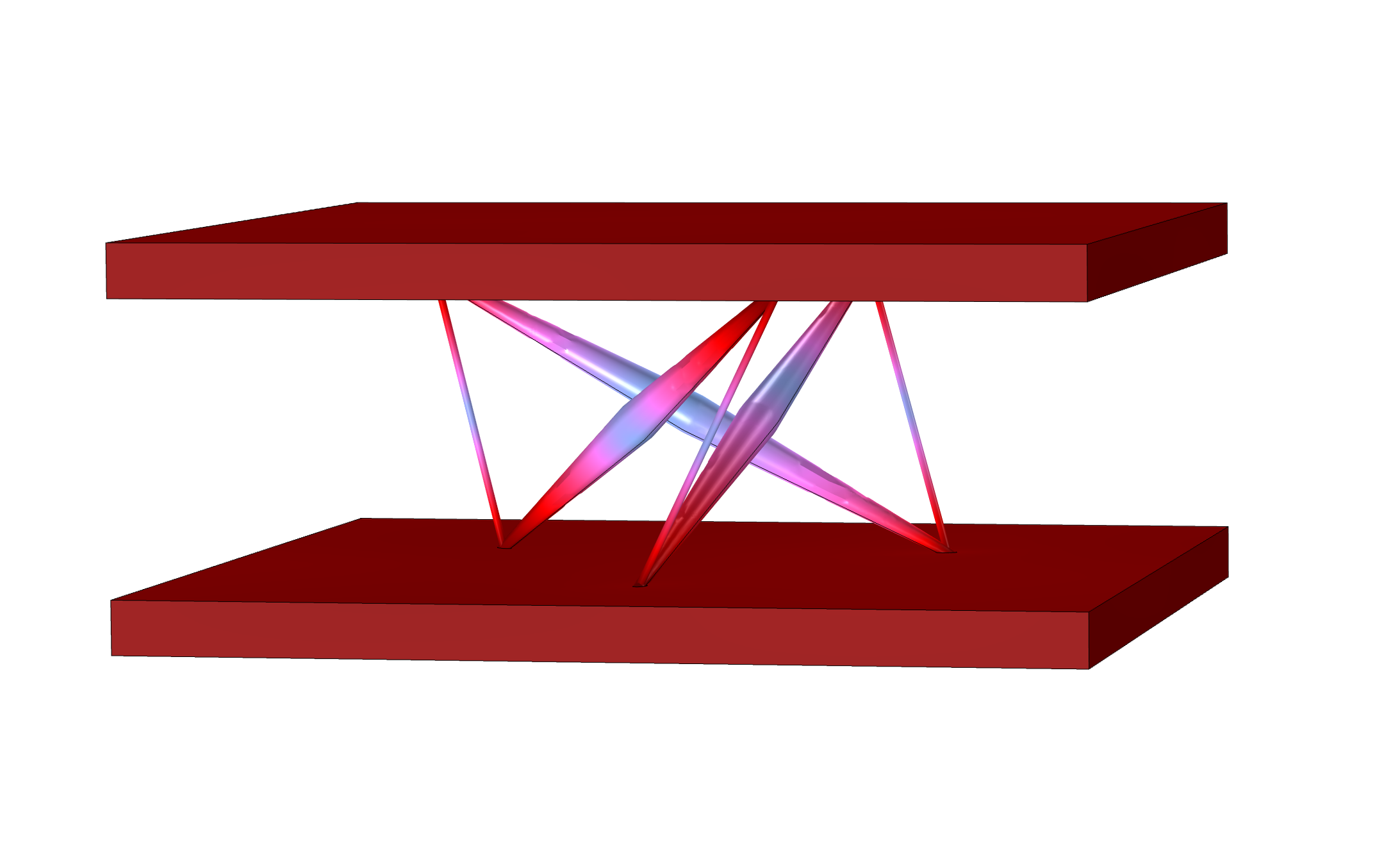}\\
				(e) $f=7.57$~kHz\\
				\includegraphics[scale=0.09]{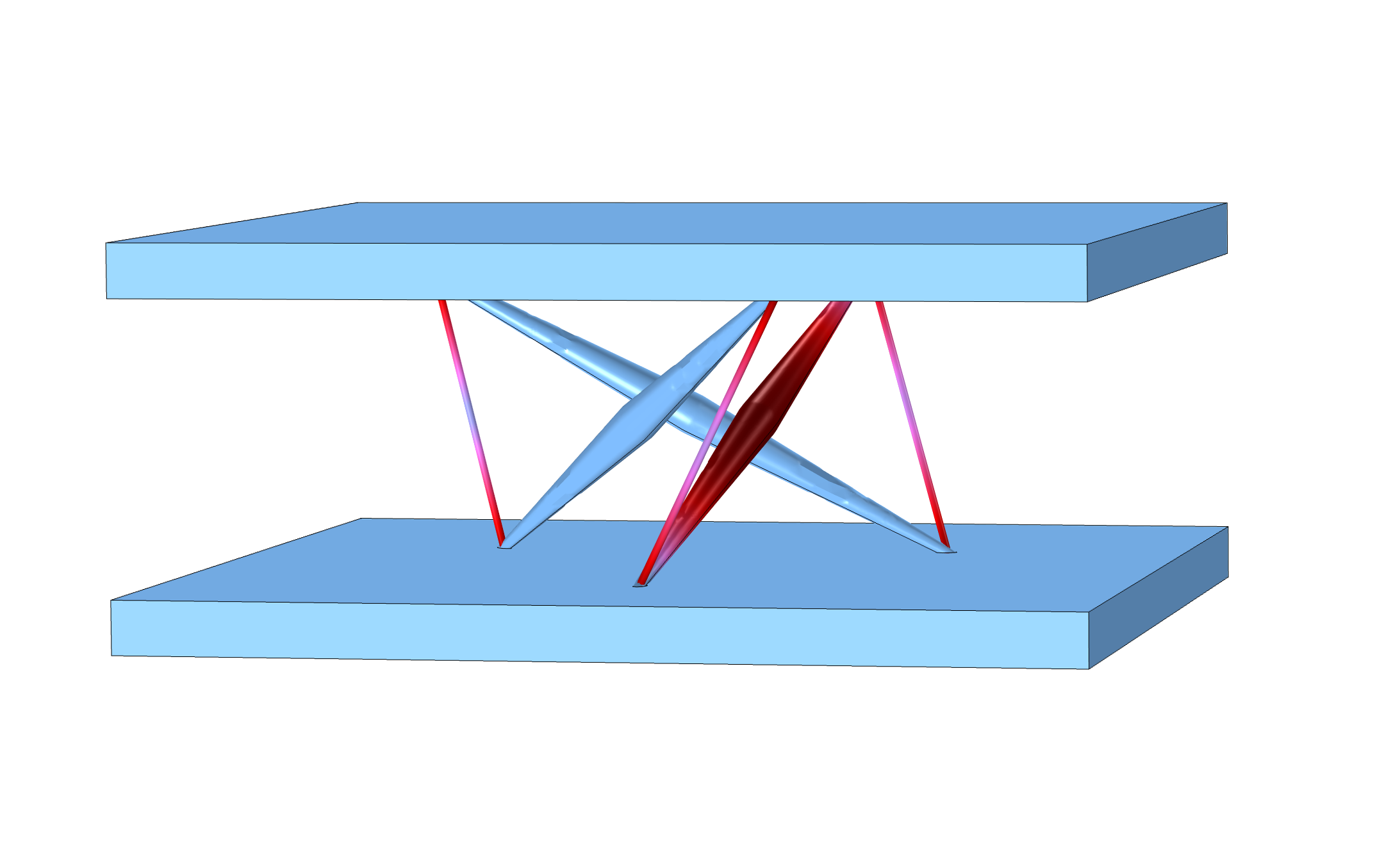}\\
				(f) $f=11.26$~kHz\\
			\end{tabular}
		\end{tabular}
	\end{tabular}
	\caption{ {\bf Elastic model with square plates}: (a) Unit cell of an accordion-like metamaterial with square terminal plates; (b) the structure of the corresponding Brillouin zone in $k$ space. (c) Dispersion relation for an accordion-like metamaterial with a representative unit cell shown in (a). Band gaps are indicated by shaded regions and numbered from 1 to 6. (d-f) Mode shapes for the 1st, 2nd and 4th modes at  Z.}
	\label{fig:squar_disk_pol}
\end{figure}

So far, we have considered wave propagation along the axial direction of a meta-chain. Stress-free boundary conditions are applied to the lateral faces of the chain resulting in a dispersion relation with multiple modes characterized by bending rotations of the discs. It is of interest to analyze the case when the excitation of bending modes is restricted. This can be achieved by considering a three-dimensional metamaterial composed by periodic replications of unit cells along three mutually perpendicular directions (see Fig.~\ref{fig:unit}b). For this purpose, we replace a circular disc by a square element of the same thickness and mass. The metamaterial unit cell is shown in Fig.~\ref{fig:squar_disk_pol}a.

The dispersion relation is evaluated numerically by applying Floquet-Bloch boundary conditions at the unit cell faces~\cite{bri46, hlr14}. The absence of translational and rotational symmetries of the unit cell requires to analyze values of wave vector ${\bf k} = \{k_x, k_y, k_z\}$ within the irreducible Brillouin zone indicated by a rectangular parallelepiped in Fig.~\ref{fig:squar_disk_pol}b. The dispersion relation dependent on three  variables cannot be graphically represented.
To simplify the analysis, we consider only specified directions and planes within the Brillouin zone, as highlighted in Fig.~\ref{fig:squar_disk_pol}b. 

%----------------------------------
%---------------------------

We first analyze the $\Gamma-Z$ direction with $k_x=0$ and $k_y=0$ corresponding to an axial wave propagation along a chain with continuity periodic boundary conditions at the lateral faces.
The dispersion relation is shown in Fig.~\ref{fig:squar_disk_pol}c. The color of  dispersion curves indicates the mode polarization from translational (blue) to bending (red). 
As can be seen, the three-dimensional accordion-like metamaterial is capable of generating multiple closely-located band gaps at low frequencies, similarly to the meta-chain with circular disks.
Comparison of the relations in Fig.~\ref{fig:squar_disk_pol}c and Fig.~\ref{fig:ref_spectrum}a reveals that the former has a smaller number of dispersion curves, most of which are translational-twisting modes.
The bending modes appear at higher frequencies, when a quarter of the wavelength of shear waves in the material approximately equals the lateral unit-cell face, i.e., $f = c_s/4h_{uc}\approx 45$~kHz with $c_s$ denoting the velocity of shear bulk waves in titanium. 
This results from the applications of periodic conditions at the lateral faces, which prohibit excitation of bending modes in tensegrity structures at lower frequencies. The mode shapes of the first translational-twisting mode (Fig.~\ref{fig:squar_disk_pol}d) resembles that of the $t2$ mode for the meta-chain (Fig.~\ref{fig:RigEl_spectrum}d). The second and third modes have very close cut-off frequencies and form a lower bound of the first band gap. The upper band-gap bound is formed by a flat curve representing a mode with  localized vibrations in a bar (Fig.~\ref{fig:squar_disk_pol}f).
The first band gap is generated at almost the same frequencies as in the meta-chain (cf. Fig.~\ref{fig:ref_spectrum}a).

%----------------------------------
\begin{figure}[h]
	\begin{tabular}{cc}
		\subfloat[]{\includegraphics[scale=0.5]{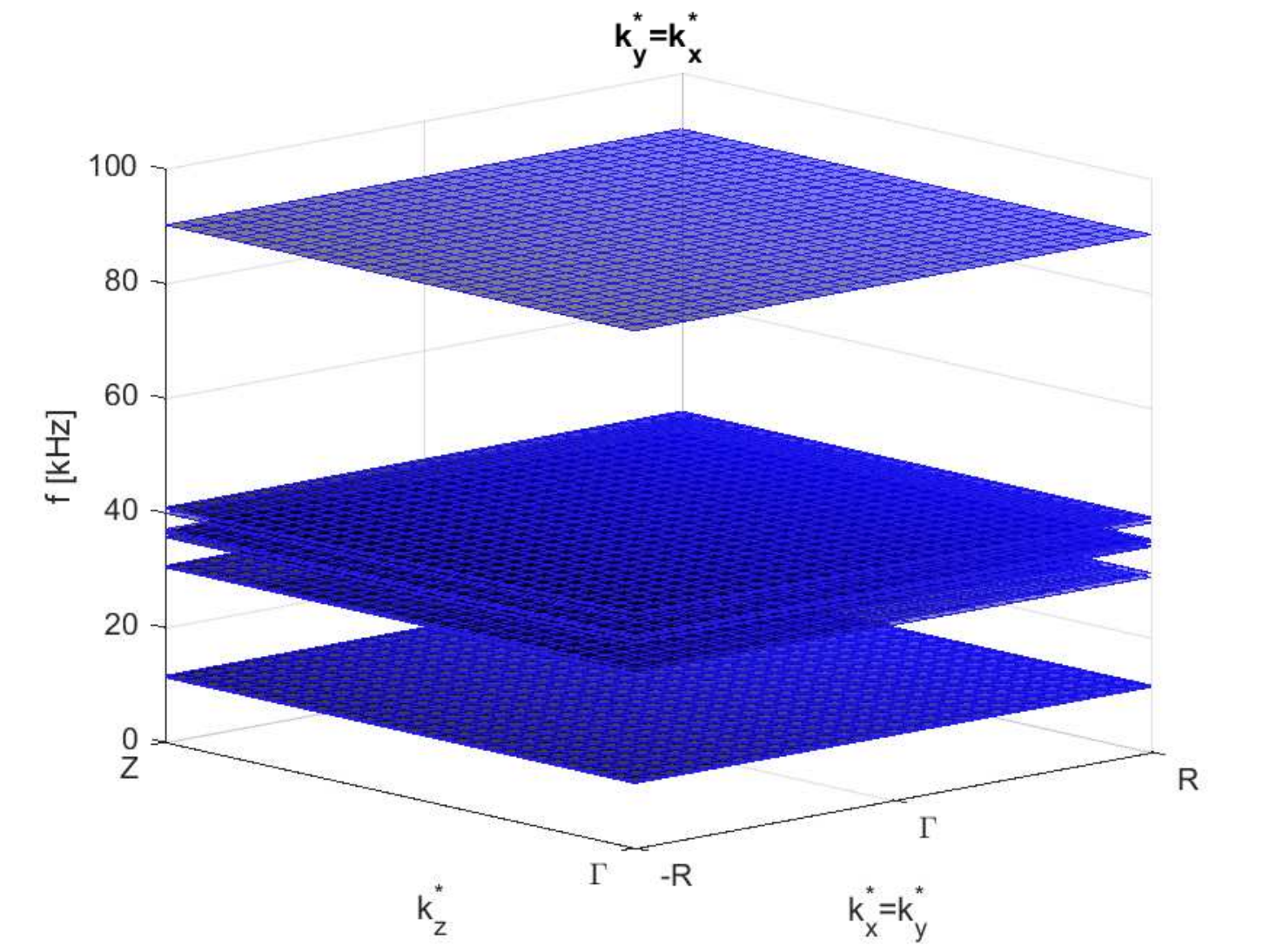}}&
		\subfloat[]{\includegraphics[scale=0.5]{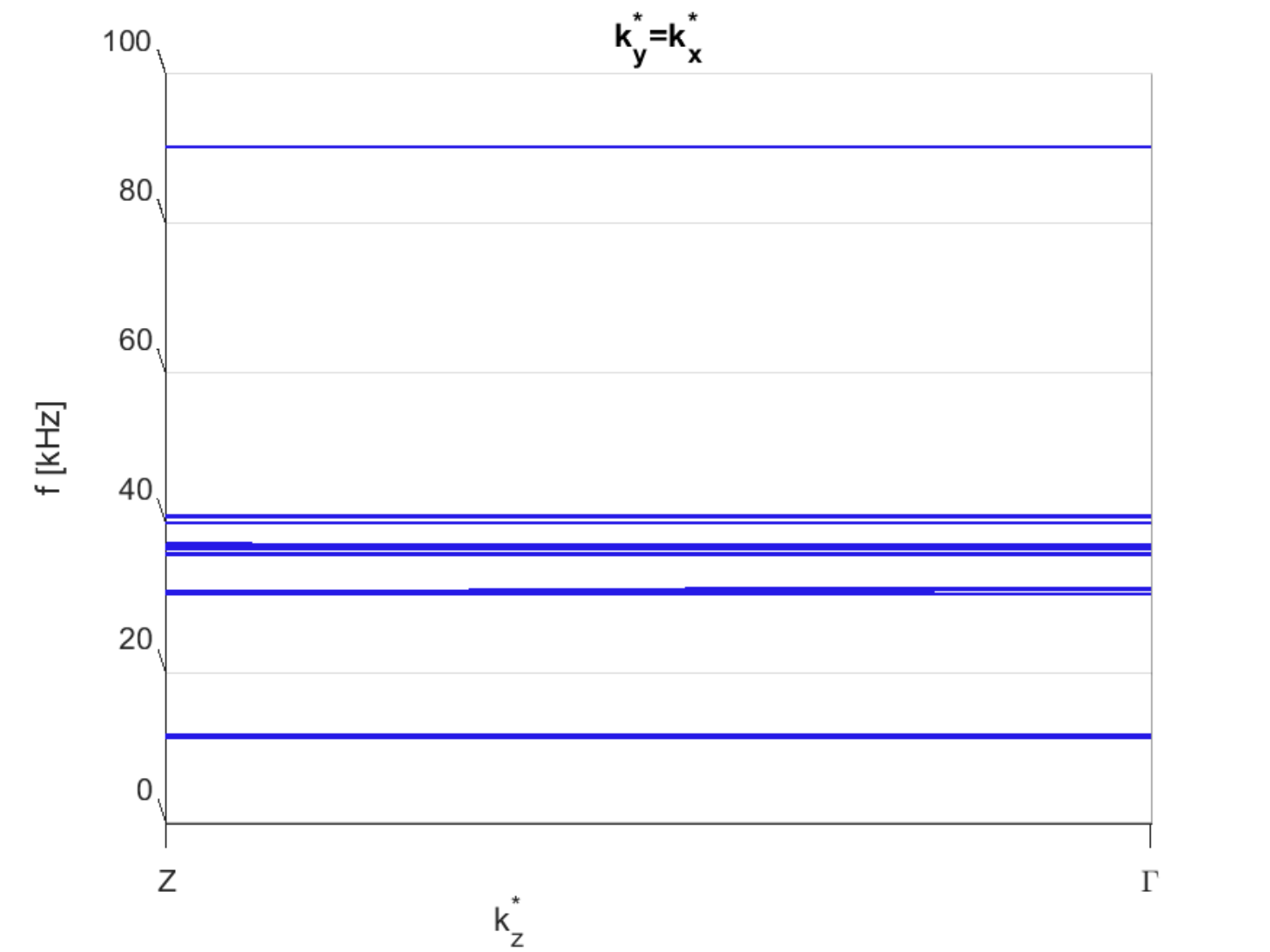}}
		\\
		\subfloat[]{\includegraphics[scale=0.5]{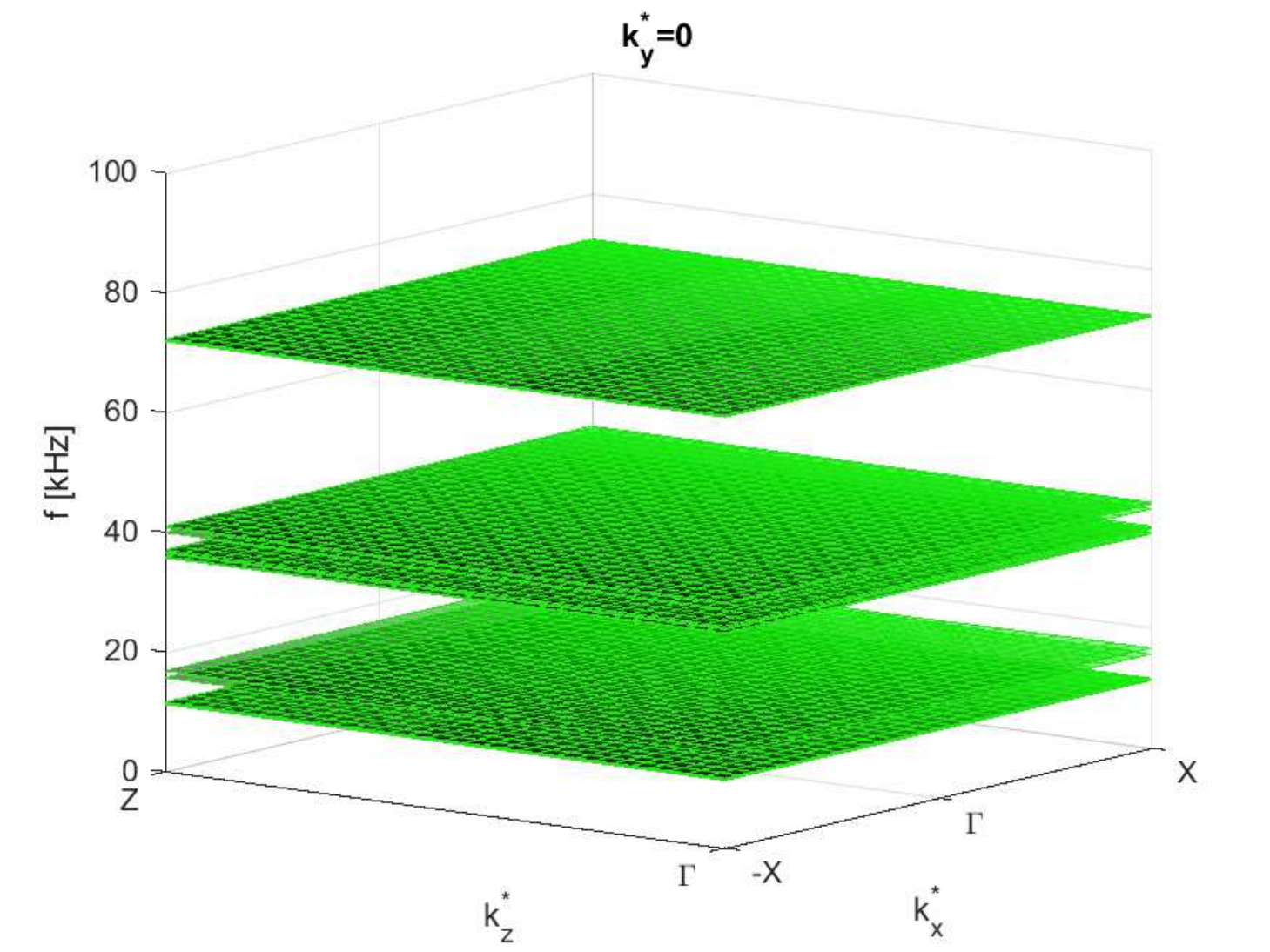}}&
		\subfloat[]{\includegraphics[scale=0.5]{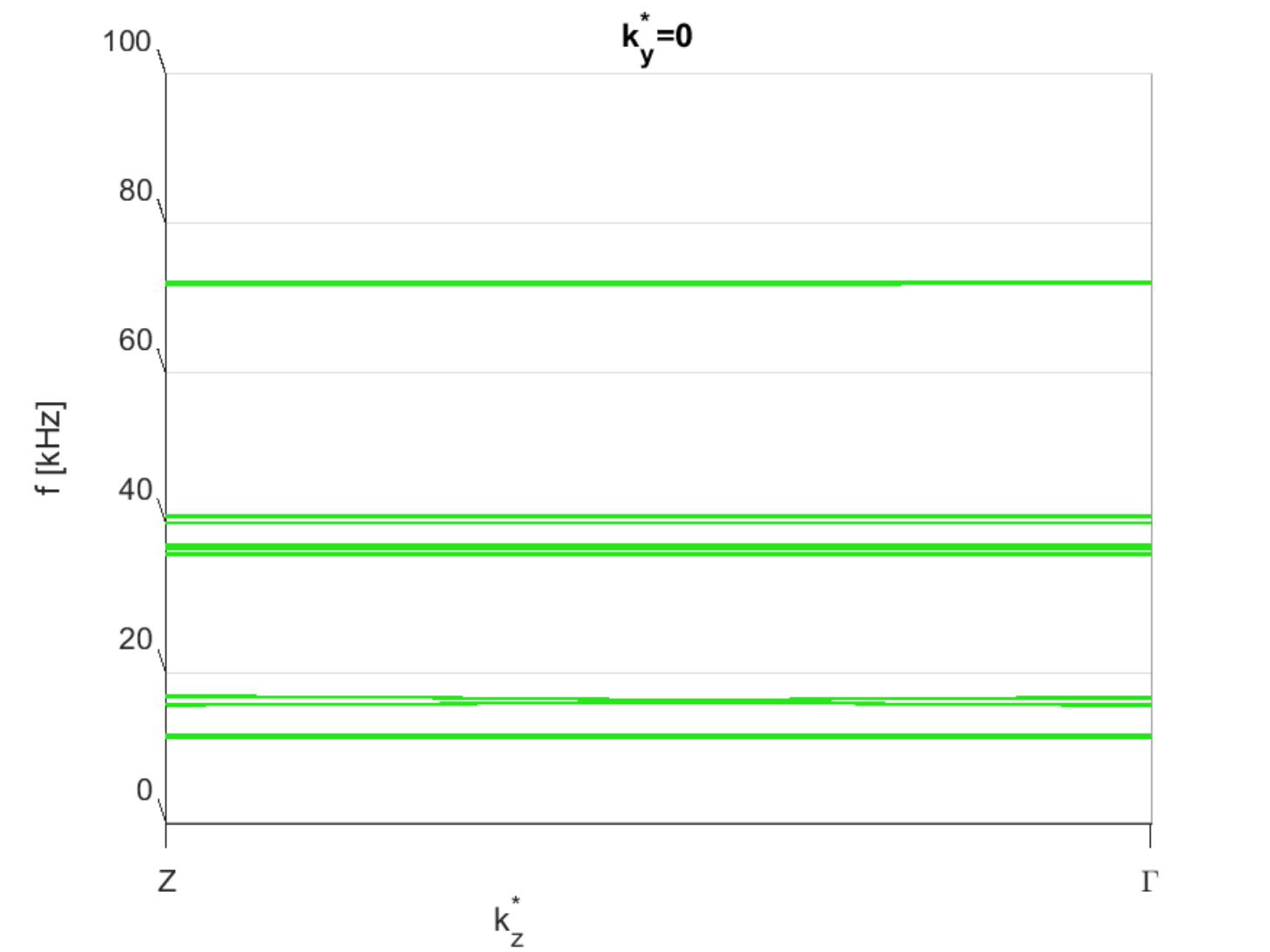}}
		\\
		\subfloat[]{\includegraphics[scale=0.5]{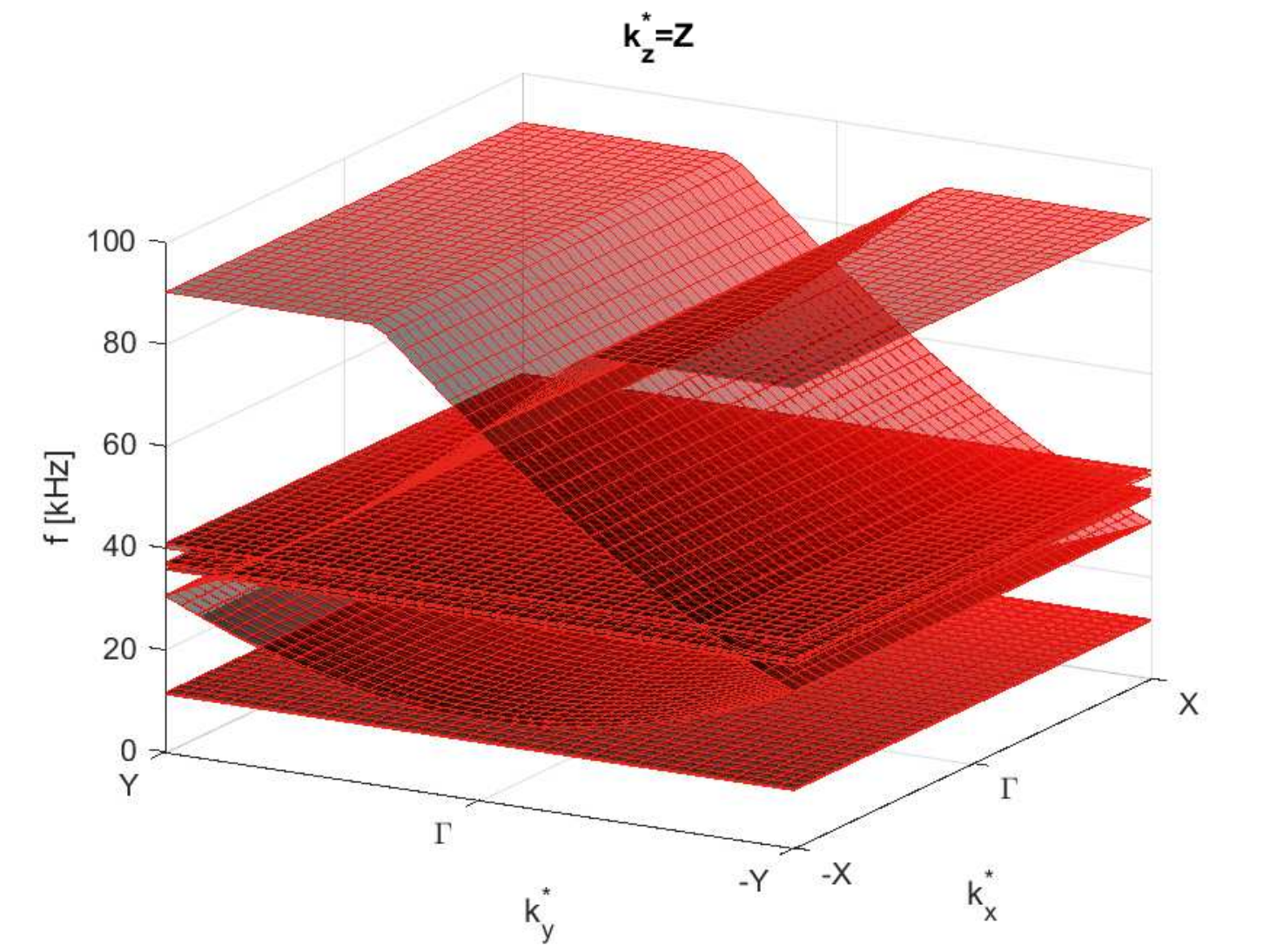}}&
		\subfloat[]{\includegraphics[scale=0.5]{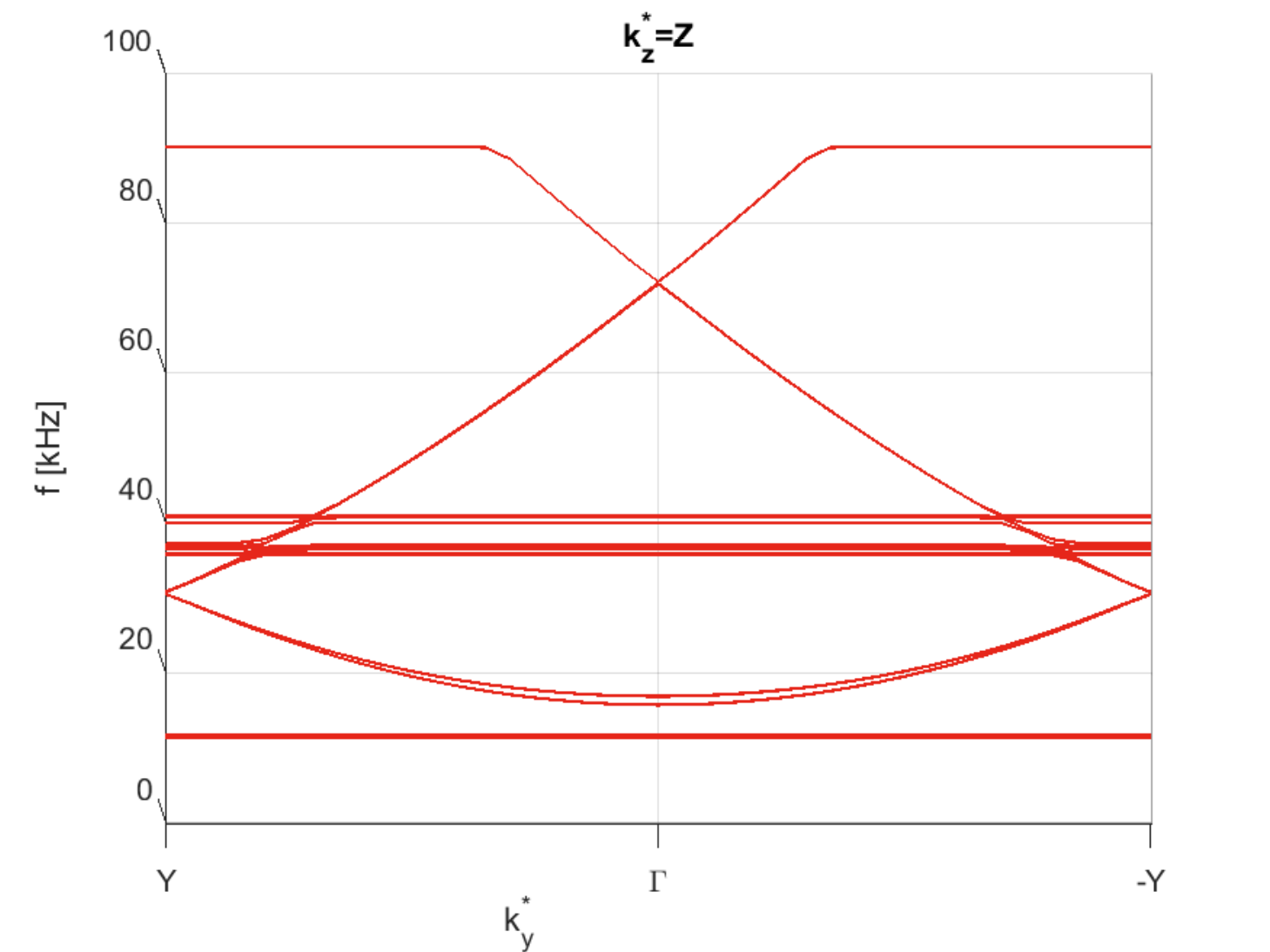}}
	\end{tabular}
	\caption{Dispersion surfaces for an accordion-like metamaterial with a representative unit cell shown in Fig.~\ref{fig:squar_disk_pol}a. Plots (b,d,f) provide two-dimensional projections of the surfaces in plots (a,c,e), respectively. The surfaces are calculated for values of the wave vector ${\bf k}=\{k_x,k_y,k_z\}$ indicated by corresponding colors in Fig.~\ref{fig:squar_disk_pol}b.}
	\label{fig:BZ_surfaces}
\end{figure}
%---------------------------

Finally, we analyze dispersion surfaces evaluated for various cross-sections of the Brillouin zone, as indicated in Fig.~\ref{fig:squar_disk_pol}b. The corresponding three-dimensional graphs and their projections are given in Fig.~\ref{fig:BZ_surfaces}. Analysis of the surfaces allows us to understand the conditions necessary to preserve band gaps for different propagation directions of incident waves. It appears that most of the band gaps are indeed preserved for oblique waves, as shown in Fig.~\ref{fig:BZ_surfaces}a-d. This makes the designed accordion-like metamaterials particularly attractive for practical applications. The worst situation occurs when waves propagate along the horizontal direction, in the plane of terminal discs (Fig.~\ref{fig:BZ_surfaces}e-f). In this case, wave attenuation mechanisms based on interactions between discs and tensegrity prisms are inefficient. However, the lowest band gap is still preserved.

\section{Conclusion}\label{sec:concl}
In summary, we have designed and studied lightweight, elastic metamaterials composed of tensegrity lattices alternated with bulky plates or discs, a structure which we have defined "accordion-like". The developed metamaterials are characterized by multiple low-frequency band gaps originating from either the Bragg scattering of modes in slender elements or a combination of Bragg scattering and local resonances. These mechanisms guarantee strong uniform wave attenuation at band-gap frequencies achieved by means of a limited number of metamaterial unit cells. The coupled band-gap generation mechanism enables to overcome a limited gap width typical for locally resonant metamaterials and provides more freedom in manipulating the band-gap frequencies by 
a selective control of locally resonant modes.
Certain adjacent band gaps can in turn be effectively merged, provided a small level of structural damping (unavoidable for real materials) is assumed.
Numerical results also show that the band gaps are resilient with respect to variations in the material properties or in the structural geometry, although optimal band-gap merger is obtained for tapered tensegrity prism bars with small end diameters.
An additional advantage of the accordion-like metamaterials is the possibility to tune band gaps by changing the level of applied prestress in the incorporated strings without introducing modifications in the metamaterial configuration.
Extending the analysis from accordion-like 1-D meta-chains to the case of 3-D arrangements, we have demonstrated that the wave attenuation properties are preserved for waves with different incidence angles.

The performed study clearly testifies the promising nature of the accordion-like designs for the realization of effective lightweight solutions for noise and vibration mitigation, acoustic cloaking, impact protection, and other applications in low-frequency ranges.
Finally, note that the features found for accordion-like metamaterials provide interesting possibilities for future extensions to`rank 2' and `rank 3' walled structures formed by structural units and confinement plates (or walls) arranged along two or three families of parallel planes~\cite{mil17}.

\section*{Acknowledgements}
A.A. gratefully acknowledges financial support from the Department of Civil Engineering at the University of Salerno. 
F.B. is supported by "Neurofibres" grant No. 732344 and by Progetto d'Ateneo/Fondazione San Paolo "Metapp", n. CSTO160004.
N.M.P. is supported by the European Commission H2020 under the Graphene Flagship Core No. 696656 (WP14 "Polymer Composites") and FET Proactive "Neurofibres" grant No. 732344.

\section*{Compliance with Ethical Standards}
The authors declare that they have no conflict of interest.

\end{document}